\documentclass[%
amsmath,aps,showkeys,
amssymb,onecolumn,pra,
superscriptaddress,
]{revtex4-2}

\usepackage{amsmath, amssymb, physics}
\usepackage{geometry}
\usepackage{braket}
\usepackage{graphicx}
\usepackage{subfigure}

\usepackage{color}
\usepackage[dvipsnames]{xcolor}
\usepackage[colorlinks=true,citecolor=NavyBlue,linkcolor=RubineRed,urlcolor=Cerulean,pdfencoding=auto]{hyperref}

\title{Effective 2D Envelope Function Theory for Silicon Quantum Dots}
\date{\today}

\begin{document}

\title{Effective 2D Envelope Function Theory for Silicon Quantum Dots}

\author{Christian W. Binder}
\affiliation{Department of Materials, University of Oxford, Parks Rd, Oxford OX1 3PJ, United Kingdom}
\affiliation{Quantum Motion, 9 Sterling Way, London N7 9HJ, United Kingdom}

\author{Guido Burkard}
\affiliation{Department of Physics, University of Konstanz, D-78457 Konstanz, Germany}

\author{Andrew J. Fisher}
\affiliation{Dept of Physics and Astronomy and London Centre for Nanotechnology,
University College London, London WC1E 6BT, United Kingdom}
\affiliation{Quantum Motion, 9 Sterling Way, London N7 9HJ, United Kingdom}

\begin{abstract}
We present a rigorous method to reduce the three-dimensional (3D) description of a quantum dot in silicon to an effective two-dimensional (2D) envelope function theory for electron spin qubits. By systematically integrating out the strongly confined vertical dimension using a Born-Oppenheimer-inspired ansatz at the envelope-function level, we derive an effective in-plane potential that faithfully captures the essential electrostatics of the full 3D system. Considering the lowest two eigenstates of the out-of-plane direction, this reduction leads to the natural and explicit emergence of the valley degree of freedom within a 2D formalism, which is derived here from first principles. We validate the accuracy of the method through comparisons with full 3D simulations and demonstrate its superiority over naive 2D slicing, particularly in the presence of interface roughness. Crucially, the reduction in dimensionality leads to substantial computational savings, making our approach particularly well suited for simulating two-electron systems, e.g., for the extraction of parameters such as the exchange coupling. Beyond its practical utility, the rigorous 2D envelope function theory that is introduced in this study incorporates valley physics in a physically grounded manner, offering conceptual clarity on the role of valley states in qubit operation and measurement.
\end{abstract}

\maketitle

\section{Introduction}

Silicon spin-based quantum computing is a promising platform for scalable quantum technologies due to its compatibility with existing CMOS infrastructure \cite{Veldhorst2014} and the long coherence times of electron spins hosted by quantum dots \cite{Kawakami2016}. The development of scalable silicon quantum devices, however, relies on fast and accurate design–simulation feedback loops \cite{veldhorst2017silicon}. As devices grow in complexity, rapid iteration capabilities over potential geometries become essential. Crucial performance metrics—such as tunnel coupling and, especially, the exchange interaction \( J \)—are highly sensitive to microscopic details \cite{yang2013} and must be computed numerically with high fidelity. Efficient yet accurate modeling tools \cite{klimeck2007_partI} are therefore indispensable for guiding experimental design and enabling the scalable control of spin qubits.

Typically, 3D simulations of the electron wavefunction are used for single-electron problems, while two-electron systems require a full 6D treatment. These computations are costly and significantly slow down the design and optimization process. Fortunately, insight can be gained by examining the shape of the electron wavefunction in a quantum dot, which—within the silicon spin qubit architecture—is realized as the spin of a single electron~\cite{Loss1998} confined near a Si/SiO$_2$ interface. The vertical confinement, imposed by the interface potential and gate-induced electric fields, is much stronger than the lateral confinement defined by surface gate electrodes. This results in a highly anisotropic, disk-like wavefunction—strongly confined in the vertical direction and extended laterally. This raises the question of whether the three-dimensional (3D) problem of a silicon-based quantum dot can be rigorously reduced to a two-dimensional (2D) model, thereby achieving significant computational savings. 2D ad hoc models have been used for a long time to describe quantum dots \cite{chakraborty1999quantum}. Other simple approaches, such as taking a fixed 2D slice through the 3D potential landscape, have been used in the literature~\cite{jnane2024},\cite{jeon2025}, but these can fail in important scenarios, particularly when surface roughness or lateral inhomogeneities strongly modulate the vertical ($z$) confinement. Hence, to model and predict the behavior of these qubits accurately, particularly for tasks such as computing tunnel couplings, exchange ($J$) couplings, and valley splittings, full three-dimensional (3D) simulations are often required. Such simulations, including mesh-based continuum models like those implemented in QTCAD~\cite{qtcad_doc,kriekouki2023} and atomistic tight-binding approaches such as NEMO3D~\cite{klimeck2007_partI,srinivasan2008}, allow for a comprehensive treatment of realistic device physics, including interface disorder. Atomistic simulations, in particular, have proven crucial for accurately capturing valley physics, a phenomenon arising from the multi-valley nature of silicon's conduction band and its interaction with interface irregularities~\cite{friesen2007}, \cite{cvitkovich2023variability}. However, these methods are computationally expensive and become unfeasible when scaled to multi-dot systems or when extensive parameter sweeps are performed.

In this study, we present a rigorous dimensional reduction scheme that projects the full three-dimensional quantum dot problem onto a two-dimensional effective model. Our method is based on the conceptual framework of the Born-Oppenheimer approximation. The central idea is to treat the out-of-plane ($z$) dynamics as fast (high-energy) compared to the slower (low-energy) in-plane ($x$-$y$) motion. This separation of scales allows us to define local  eigenfunctions and eigenenergies for the $z$ motion, which paramaterically depend on $x$ and $y$ and serve as inputs for a projected two-dimensional Hamiltonian.
If we retain only the lowest eigenstate of the vertical Hamiltonian, the resulting model describes purely orbital dynamics in the 2D plane. In the orbital regime, our method rigorously incorporates the effects of interface roughness and spatially varying vertical confinement across both single and double quantum dot systems, enabling accurate predictions of quantities such as tunnel and exchange ($J$) couplings within a two-dimensional framework.

Moreover, by including the two lowest eigenstates, which typically correspond to the lowest pair of valley states, we naturally recover also the valley degree of freedom (DOF). This statement is usually justified by the large energy gap to higher orbital and additional valley states, which are typically lifted by strain and interface effects in silicon quantum dots \cite{yang2013spin}.
This valley structure originates from the near-degenerate conduction band minima along the $\pm z$ directions in the silicon bandstructure~\cite{friesen2007} and is essential for modeling realistic quantum dot qubits.

The resulting 2D theory includes spatially dependent coupling matrices that govern inter-valley interactions, thus providing a tool to model the orbit and valley physics of quantum dots. While spin could, in principle, be incorporated via effective, dot-specific $g$-factors, this issue is not addressed in the present work. 

Our method also offers clear didactic value through the natural emergence of a two-component valley wavefunction. This feature provides a transparent and computationally accessible framework for analyzing how valley physics influences the qubit measurement and decoherence processes, as well as two-qubit gate operations.
To validate our approach, we compare results from the 2D projected model with full 3D simulations - both performed using grid-based tools. More specifically, we benchmark our methods against naive 2D slicing approaches and demonstrate their superior accuracy in computing tunnel and exchange couplings, as well as valley splittings and phases.

\section{Theory}

The aim of this section is to derive effective 2D Hamiltonians that accurately capture the physics of quantum dots by systematically reducing the full 3D problem. The central strategy draws inspiration from the Born-Oppenheimer approximation~\cite{bornoppenheimer1927}, where a system is separated into fast and slow degrees of freedom. In our context, the strong vertical confinement along the \( z \)-direction produces fast dynamics, while the slower in-plane motion occurs along the \( x \) and \( y \) directions.

\subsection{Modeling Framework}

We begin with a brief review of how to describe the quantum dot wavefunction in a gate-defined silicon quantum dot, along with existing methods to simplify its representation. In general, the wavefunction of an electron in a silicon heterostructure experiences a highly non-trivial potential consisting of three components: the silicon crystal potential, an abrupt dielectric interface between silicon and the barrier material (e.g., silicon-oxide), and the external electrostatic potential generated by gate electrodes. Hence, the total Hamiltonian describing the system in 3$D$ can be written as:
\begin{equation}
H = \underbrace{-\frac{\hbar^2}{2m_e} \nabla^2 + V_{\text{crystal}}(\mathbf{r})}_{H_{\text{crystal}}} + V(\mathbf{r}),
\label{eq:total_H}
\end{equation}
where \( H_{\text{crystal}} \) is the Hamiltonian of the unperturbed silicon crystal and \( V(\mathbf{r}) \) includes all additional contributions from the dielectric interface and the gate-induced electrostatic potential.
The crystal and interface potentials vary on atomistic length scales, imposing formidable spatial resolution requirements. Furthermore, only conduction-band states are relevant, as the valence band is fully occupied under typical conditions. Consequently, a proper description of the electron state involves a superposition of Bloch states from the lowest conduction-bandeigenstates of \(H_{\text{crystal}}\). Using these Bloch states as a basis, the single-particle Schr\"odinger equation becomes:
\begin{equation}
\sum_{\mathbf{k}} \left[ \epsilon(\mathbf{k}) + V(\mathbf{r}) \right] \phi_{\mathbf{k}}(\mathbf{r}) c_\mathbf{k} = E \sum_{\mathbf{k}} \phi_{\mathbf{k}}(\mathbf{r}) c_\mathbf{k},
\label{eq:bloch_basis}
\end{equation}
where \(\phi_{\mathbf{k}}(\mathbf{r}) = u_{\mathbf{k}}(\mathbf{r}) e^{i\mathbf{k} \cdot \mathbf{r}}\) are Bloch functions satisfying $H_{\text{crystal}}\phi_{\mathbf{k}}=\epsilon(\mathbf{k})\phi_{\mathbf{k}}$, and \(\Psi(\mathbf{r}) = \sum_{\mathbf{k}} \phi_{\mathbf{k}}(\mathbf{r}) c_{\mathbf{k}}\) is the total wavefunction. 
Here, \(\epsilon(\mathbf{k})\) denotes the energy dispersion of the lowest conduction band. Simulation tools such as QTCAD adopt this full-band Bloch-state representation~\cite{qtcad_doc}.

While accurate, this formulation is computationally demanding due to the fast spatial oscillations in the wavefunction. These arise from two sources:

\begin{itemize}
    \item The periodic functions \(u_{\mathbf{k}}(\mathbf{r})\) vary on atomistic length scales, typically much smaller than the lattice constant. For silicon, the lattice constant is \(a_{\text{Si}} = 0.543~\text{nm}\).
    \item Silicon’s conduction band features six equivalent minima (valleys) located along the \(\langle 100 \rangle\) directions in the Brillouin zone. Under confinement and strain, the four in-plane valleys (\(\pm \hat{x}, \pm \hat{y}\)) are typically raised in energy (see, e.g., \cite{friesen2006magnetic}), leaving two low-energy valleys along the \(\pm \hat{z}\) direction. These lead to interference patterns in the wavefunction on the scale of \(1/|\mathbf{k}_0|\), where \(\pm \mathbf{k}_0 = \pm 0.85 \times \frac{2\pi}{a_{\text{Si}}} \hat{z} \approx \pm 9.83~\text{nm}^{-1} \hat{z}\) are the valley wavevectors.
\end{itemize}

Due to this structure, the quantum dot ground state is often approximated as
\begin{equation}
\label{eq:envappr}
\Psi(\mathbf{r}) = \sum_{\mathbf{k}} u_{\mathbf{k}}(\mathbf{r}) e^{i \mathbf{k}  \cdot \mathbf{r}} c_{\mathbf{k}} 
\approx F_+(\mathbf{r}) u_{+\mathbf{k}_0}(\mathbf{r}) e^{i \mathbf{k}_0 \cdot \mathbf{r}} 
+ F_-(\mathbf{r}) u_{-\mathbf{k}_0}(\mathbf{r}) e^{-i \mathbf{k}_0 \cdot \mathbf{r}},
\end{equation}
Here, \(F_\pm\) is defined as
\begin{equation}
F_\pm(\mathbf{r}) = \sum_{|\mathbf{q}| \ll \frac{1}{a} } c_{\pm \mathbf{k}_0 + \mathbf{q}} e^{i \mathbf{q} \cdot \mathbf{r}}.
\end{equation}

This approximation assumes that only states near the valley minima contribute. This is equivalent to saying that $F_\pm(\mathbf{r})$ does not vary much on the scale of a unit cell. Furthermore, in \eqref{eq:envappr} we used,
\begin{equation}
u_{\mathbf{k}}(\mathbf{r}) \approx u_{\pm \mathbf{k}_0}(\mathbf{r}) \quad \text{for } \mathbf{k} \approx \pm \mathbf{k}_0,
\end{equation}
effectively separating the wavefunction into slowly varying envelopes, \(F_\pm(\mathbf{r})\), and rapidly oscillating phase factors. This approximation is usually called \textit{envelope function approximation}~\cite{yu2010fundamentals,Friesen2013}.

In scenarios where fine details of valley physics are not essential—such as estimating exchange couplings \(J\) or simulating charge stability diagrams—a single-valley approximation is typically sufficient. In this case, it is assumed that the envelope functions satisfy \( |F_+(\mathbf{r})| = |F_-(\mathbf{r})| = |F(\mathbf{r})| \), i.e., they are equal in magnitude but may differ by a phase. The resulting envelope satisfies an effective-mass Schrödinger equation~\cite{luque2015single}.

However, when valley-specific effects become relevant, intervalley coupling or valley-orbit coupling must be retained. These couplings arise from high-frequency Fourier components of the external potential, particularly those near the wavevector \( 2\mathbf{k}_0 \), via terms like \( \tilde{V}(2\mathbf{k}_0) \), which mediate intervalley scattering. Here, \( \tilde{V}(\mathbf{k}) \) denotes the Fourier transform of \(V(\mathbf{r}) \).

\vspace{2em}
\textbf{Generalized Envelope Function (GEF) Approach}
\vspace{1em}

We now propose an intermediate model between the full Bloch-state approach and the envelope approximation. In this \textit{Generalized Envelope Function} (GEF) framework, we remove the periodic part $u_{\pm \mathbf{k}_0}(\mathbf{r})$ of the  Bloch functions while retaining the essential fast oscillations arising from the multi-valley structure. All Bloch-function complexity is abstracted into a single momentum-dependent kernel.
We define the GEF as the spatial representation of the $k$-space coefficients. In a finite normalization volume $\Omega$,
\begin{equation}
\psi(\mathbf r)=\frac{1}{\sqrt{\Omega}}\sum_{\mathbf k} c_{\mathbf k}\,e^{i\mathbf k\cdot\mathbf r},
\label{eq:gef_discrete}
\end{equation}
and, in the continuum limit,
\begin{equation}
\psi(\mathbf r)=\int \frac{d^3k}{(2\pi)^{3/2}}\,c(\mathbf k)\,e^{i\mathbf k\cdot\mathbf r}.
\label{eq:gef_continuum}
\end{equation}
Comparing with the full wavefunction
\(
\Psi(\mathbf r)=\sum_{\mathbf k} u_{\mathbf k}(\mathbf r)\,e^{i\mathbf k\cdot\mathbf r}\,c_{\mathbf k},
\)
the GEF removes only the atomically varying factors $u_{\mathbf k}(\mathbf r)$ while keeping the valley-phase oscillations carried by $e^{i\mathbf k\cdot\mathbf r}$.  

When two valleys at $\pm\mathbf k_0$ dominate, $\psi$ admits the standard two-valley decomposition with slowly varying envelopes,
\begin{equation}
\psi(\mathbf r)\;\approx\;F_+(\mathbf r)\,e^{+i\mathbf k_0\cdot\mathbf r}
\;+\;F_-(\mathbf r)\,e^{-i\mathbf k_0\cdot\mathbf r},
\label{eq:env_function_theory}
\end{equation}
so that, upon restriction to the valley subspace, the GEF reduces to the familiar envelope-function model with valley-dependent envelopes $F_\pm(\mathbf r)$.
The governing equation for the GEF is an effective Schrödinger equation,
\begin{equation}
\left( \mathcal{F} \, \epsilon(\mathbf{k}) \, \mathcal{F}^{-1} + V_{\text{eff}}(\mathbf{r}) \right) \psi(\mathbf{r}) = E \, \psi(\mathbf{r}),
\label{eq:gef_schrodinger}
\end{equation}
where \(\mathcal{F}\) denotes the Fourier transform operator. The effective potential \(V_{\text{eff}}(\mathbf{r})\) is given by:
\begin{equation}
V_{\text{eff}}(\mathbf{r}) = \int \frac{d^3q}{(2\pi)^{3/2}} C_0(\mathbf{q}) \tilde{V}(\mathbf{q}) e^{i\mathbf{q} \cdot \mathbf{r}},
\label{eq:veff}
\end{equation}
where \(\tilde{V}(\mathbf{q})\) is the Fourier transform of the external potential and \(C_0(\mathbf{q})\) is an overlap kernel that captures the effect of the underlying Bloch functions. A detailed derivation of Eq.~\eqref{eq:gef_schrodinger} and~\eqref{eq:veff} and the structure of \(C_0(\mathbf{q})\) is presented in the Appendix~\ref{app:app_der_mpf}.

The generalized envelope function forms the core element of the models and simulations presented in this work. One major advantage of this formulation is the ability to seamlessly transform between real and momentum space via Fourier transforms, enabling efficient computation through the use of fast Fourier transform (FFT) algorithms (Sec.~\ref{sec:comp_imp}). The encapsulation of the Bloch function contributions within the factor \( C_0 \) is motivated by prior studies on multivalley envelope function theory in silicon systems~\cite{thayil2024theory, hosseinkhani2020electromagnetic, saraiva2011intervalley}.

As a model that should capture all relevant features of the bandstructure, we choose to approximate \(\epsilon(\mathbf{k})\) as a sum of two anisotropic Gaussians centered at the valley minima near \(\pm k_0 \hat{z}\):
\begin{equation}
\epsilon(\mathbf{k}) = -\mathcal{H} \left[ e^{-(\mathbf{k} + k_0 \hat{z})^\top \kappa (\mathbf{k} + k_0 \hat{z})} + e^{-(\mathbf{k} - k_0 \hat{z})^\top \kappa (\mathbf{k} - k_0 \hat{z})} \right],
\label{eq:eff_bandstructure}
\end{equation}
where \(\kappa\) is a diagonal inverse effective mass tensor with components \(\kappa_{xx} = \kappa_{yy} = \hbar^2 / (2 m_t \mathcal{H})\) and \(\kappa_{zz} = \hbar^2 / (2 m_z \mathcal{H})\). Here, \(m_t\) and \(m_z\) denote the transverse and longitudinal effective masses, \(\mathcal{H} = 1000\,\text{meV}\) is a phenomenological energy scale, and the magnitude of the valley wavevector is \(k_0 = 9.83\,\text{nm}^{-1}\). This bandstructure reproduces the correct effective masses—defined as the curvature of the conduction band—at the valley minima, thereby generalizing the standard effective mass approximation of silicon to a multivalley formulation~\cite{friesen2007,friesen2006magnetic}. The use of anisotropic Gaussians is motivated not only by their ability to reproduce the local effective masses, but also by the fact that they smoothly interpolate between the two valleys and decay toward the Brillouin zone center. This provides a more realistic description of the band dispersion near the \(\Gamma\) point compared to a piecewise parabolic model, which may introduce artificial discontinuities.

For non-valley-resolved calculations, we only consider a single minimum of the bandstructure and may therefore resort to a simpler representation of the kinetic energy operator, valid within the effective mass approximation. In this case, the kinetic energy operator exhibits the standard form
\begin{equation}
\hat{\mathcal{T}} = -\frac{\hbar^2}{2 m_t} \left( \frac{\partial^2}{\partial x^2} + \frac{\partial^2}{\partial y^2} \right) - \frac{\hbar^2}{2 m_z} \frac{\partial^2}{\partial z^2},
\label{eq:charge_ham}
\end{equation}

which is justified as here, we approximate the conduction band dispersion at the valley minimum harmonically and can therefore identify the kinetic term with differential operators.

\subsection{Born-Oppenheimer-Inspired Dimensional Reduction}

In this section, we outline our procedure to perform an effective dimensional reduction of the quantum dot Hamiltonian. There are two central requirements we have on a Hamiltonian in order to apply our approach. The first is that the full Hamiltonian \( \hat{H} \) is approximately separable into a vertical and an in-plane component:
\begin{equation}
    \hat{H}(x,y,z) \approx \hat{H}_{xy}(x, y) + \hat{H}_z(z ; x, y).
\label{ass:separableHamiltonian}
\end{equation}
Here, \( \hat{H}_z \) must only exhibit parametric dependence on \( x \) and \( y \); in particular, no \( x \) or \( y \) derivatives should appear in \( \hat{H}_z \). Since this separability requirement is trivial for the potential, it translates to the requirement that the (usually non-diagonal) position representation of the kinetic energy operator is separable, i.e., \( \hat{T} = \hat{T}_{xy}(x, y) + \hat{T}_z(z; x, y) \).
The second requirement is that the kinetic energy operator in the horizontal Hamiltonian $H_{xy}$ is, at least approximately, given by
\begin{equation}
\label{ass:kinetic_energy_operator}
\hat{T}_{xy} = -\frac{\hbar^2}{2m_t} (\partial_x^2 + \partial_y^2). 
\end{equation}
The validity of these approximations in our cases of interest will be addressed in later sections.

Within this framework, we are now attempting to derive an effective two-dimensional Hamiltonian. We begin by expressing the full wavefunction as an expansion in the local eigenstates \( \chi_m(x, y, z) \) of \( \hat{H}_z \),
\begin{equation}
\psi(x, y, z) = \sum_m \phi_m(x, y)\, \chi_m(x, y, z),
\end{equation}
where \( \phi_m(x, y) \) are slowly varying envelope functions. Inserting this ansatz into the full, time-independent Schrödinger equation \( \hat{H}(x,y,z) \psi(x,y,z) = E \psi(x,y,z) \), multiplying on the left by \( \chi^*_{m'}(x, y, z) \), and integrating over \( z \), we obtain a coupled set of 2D equations:
\begin{equation}
\sum_m \hat{H}^{\text{eff}}_{m'm}(x, y)\, \phi_m(x, y) = E\, \phi_{m'}(x, y),
\end{equation}
where the matrix elements of the effective Hamiltonian are given by
\begin{align}
\hat{H}^{\text{eff}}_{m'm}(x, y) &= \int dz\, \chi^*_{m'}(x, y, z)\, \hat{H}(x, y, z)\, \chi_m(x, y, z) \notag \\
&= \left( -\frac{\hbar^2}{2m_t} (\partial_x^2 + \partial_y^2) + V_{xy}(x, y) + \epsilon_m(x, y) \right) \delta_{m'm} \notag \\
&\quad - \frac{\hbar^2}{2m_t} \left[
D^{(0)}_{m'm}(x, y)
+ D^{(1,x)}_{m'm}(x, y)\, \partial_x
+ D^{(1,y)}_{m'm}(x, y)\, \partial_y
\right],
\label{eq:H_eff_final}
\end{align}
with coupling coefficients
\begin{equation}
\begin{aligned}
D^{(0)}_{m'm}(x, y) &= \int dz\, \chi^*_{m'}\, (\partial_x^2 + \partial_y^2) \chi_m, \\
D^{(1,x)}_{m'm}(x, y) &= 2 \int dz\, \chi^*_{m'}\, \partial_x \chi_m, \\
D^{(1,y)}_{m'm}(x, y) &= 2 \int dz\, \chi^*_{m'}\, \partial_y \chi_m.
\end{aligned}
\label{eq:coupling_coeffs}
\end{equation}

This framework is closely related to the Born–Oppenheimer approximation in quantum chemistry, where the wavefunction is separated into fast electronic and slow nuclear components. Solving the electronic problem first yields potential energy surfaces (PES) for the nuclear motion—formally analogous to the eigenenergies \( \epsilon_m(x, y) \) of the vertical Hamiltonian in our case.
If all vertical basis states are retained, the expansion is exact and yields an infinite set of coupled 2D equations—just like the full coupled PES picture in molecular systems. However, practical implementations require truncation. Often, a single surface (the ground-state PES) suffices, especially when non-adiabatic couplings are small~\cite{tully1998, bornoppenheimer1927}. In our setting, this corresponds to keeping only the basis state corresponding to the lowest vertical eigenenergies,  \( \chi_0(x,y,z) \), yielding a 2D charge model. Including the two lowest vertical states—typically a near-degenerate valley pair—introduces valley-valley coupling in the effective theory, analogous to non-adiabatic transitions between PES in molecules.
This projection framework provides a simple and efficient method for dimensional reduction. In the following sections, we derive explicit forms of the resulting 2D Hamiltonians in both the charge-only and valley-coupled regimes.

\subsection{Effective 2D Hamiltonian in Charge Space}

We now apply the projection framework to the simplest physically relevant case: retaining only the lowest vertical eigenstate \( \chi_0(x, y, z) \). This corresponds to a single-surface approximation, where valley effects are neglected and only charge dynamics are captured.
Starting from the full 3D Hamiltonian for a single electron in a quantum dot illustrated in Eq.~\eqref{eq:charge_ham}, with vertical contribution
\begin{equation}
\hat{H}_z(x, y, z) = -\frac{\hbar^2}{2m_z}\partial_z^2 + V_z(x, y, z),
\end{equation}
we note that the decomposition of the total potential \( V(x, y, z) \) into an in-plane component \( V_{xy}(x, y) \) and a vertical component \( V_z(x, y, z) \) is not unique. In fact, there are infinitely many ways to write \( V(x, y, z) = V_{xy}(x, y) + V_z(x, y, z) \). A natural and physically motivated choice is to define \( V_{xy}(x, y) \) by slicing the 3D potential at a fixed vertical position \( z = z' \), such that
\begin{equation}
V_{xy}(x, y) := V(x, y, z = z'), \quad \text{and} \quad V_z(x, y, z) := V(x, y, z) - V_{xy}(x, y).
\end{equation}

As a special case of Eq.~\eqref{eq:H_eff_final}, retaining only the ground vertical state \( m = 0 \) yields a particularly simple effective Hamiltonian. In this single-surface approximation, the first-order coupling terms in Eq.~\eqref{eq:coupling_coeffs} vanish identically, and we neglect the second-order correction \( D^{(0)}_{00}(x, y) \). The integration in Eq.~\eqref{eq:H_eff_final} contributes a local energy shift, which we denote \( \epsilon_0(x, y) \), corresponding to the ground-state energy of the vertical Hamiltonian \( \hat{H}_z(x, y, z) \). The resulting 2D Hamiltonian is
\begin{equation}
\hat{H}_{\text{eff}}(x, y) = -\frac{\hbar^2}{2m_t}(\partial_x^2 + \partial_y^2) 
+ \underbrace{V_{xy}(x, y) + \epsilon_0(x, y)}_{=: V_{\text{eff}}(x, y)},
\label{eq:effective_charge_ham}
\end{equation}
where the new effective 2$D$ potential, $V_{\text{eff}}(x, y)$, is simply the sum of the in-plane potential \( V_{xy}(x, y) \) and the vertical ground-state energy \( \epsilon_0(x, y) \).
Equation~\eqref{eq:effective_charge_ham} defines a compact and computationally efficient 2D model for describing charge dynamics in quantum dots, with valley effects neglected by construction.

\subsection[Effective 2D Multivalley Envelope Function Theory]{Effective Hamiltonian in Charge and Valley Space — 2D Multivalley Envelope Function Theory}

In this subsection, we evaluate the projected Hamiltonian \eqref{eq:H_eff_final} using the two lowest vertical eigenstates, which—as discussed earlier— usually form a near-degenerate valley pair. 
To be permitted to use the general form of Eq.~\eqref{eq:H_eff_final}, we must first assess whether the Hamiltonian is approximately separable as assumed in Eq.~\eqref{ass:separableHamiltonian} and whether the form of the kinetic energy operator is in accordance with Eq.~\eqref{ass:kinetic_energy_operator}. Starting from the full Hamiltonian \eqref{eq:H_eff_final}, and using the anisotropic bandstructure model of Eq.~\eqref{eq:eff_bandstructure}, we proceed by making a further simplification: we approximate the in-plane (\( x \)-\( y \)) bandstructure quadratically at low energies, i.e., we assume that the bandstructure can be written as a parabolic function in $x$ and $y$, superposed with the double well function in $z$. Under these assumptions, the Hamiltonian is approximately separable, and we are justified in applying the dimensional reduction framework discussed in the previous section.

A complication, however, arises from the fact that the valley structure of the vertical eigenstates is highly sensitive to interface disorder, especially in realistic Si/SiGe or Si/SiO\(_2\) devices.  
To avoid spatial fluctuations in the basis, and thereby suppressing the coupling terms in Eq.~\eqref{eq:coupling_coeffs}, we choose to work in a basis where we fix the valley-dependent terms in the vertical eigenfunctions (but still allow the vertical envelope function to vary as a function of $x$ and $y$); we call this the `fixed valley-spinor basis' \( \{ \tilde{\chi}_+, \tilde{\chi}_- \} \) (see Appendix~\ref{app:ValleyHamiltonian} for details of how it is constructed). As this basis is different from the eigenbasis of the local vertical Hamiltonian $\hat{H}_z$ we are required to express the diagonal $\epsilon_m(x, y) \delta_{m'm}$ term in \eqref{eq:H_eff_final} in it. We call this rotated representation of $\epsilon_m(x, y) \delta_{m'm}$, $\hat{H}_z^{(+,-)}(x,y)_{\nu' \nu}$. In this fixed valley basis, the coupling terms \eqref{eq:coupling_coeffs} manifest only in the $xy$-variation of the envelope function. Provided that coupling to higher energy states can be neglected, the effective 2D multivalley Hamiltonian becomes
\begin{equation}
\hat{H}^{\text{eff}}_{\nu' \nu}(x,y) =
\left( -\frac{\hbar^2}{2m_t} (\partial_x^2 + \partial_y^2) + V_{xy}(x,y) \right) \delta_{\nu' \nu}
+ \hat{H}_z^{(+,-)}(x,y)_{\nu' \nu},
\label{eq:matvalH}
\end{equation}
where \( \nu, \nu' \in \{ +, - \} \) label the valley-spinor components.
The valley-dependent part of the Hamiltonian \( \hat{H}_z^{\text{valley}}(x,y) \) captures local valley splitting and mixing,
\begin{equation}
\label{eq:matvalHv}
\hat{H}_z^{(+,-)}(x,y) =
\epsilon_g(x,y)+2|\Delta|(x,y)\begin{pmatrix}
\cos^2\phi & -\cos\phi\sin\phi\\
-\cos\phi\sin\phi & \sin^2\phi
\end{pmatrix},
\end{equation}
where \( \epsilon_g(x,y) \) is the local valley ground state energy and \( |\Delta(x,y)| \) is the position-dependent valley splitting. Finally, \( \phi(x,y) =\arg \Delta(x,y)\) denotes the local valley phase, defined through the complex phase of the intervalley coupling. 
The matrix in the second term of Eq.~\eqref{eq:matvalHv} captures the essential valley physics emerging from interface roughness and confinement-induced mixing. Both Eq.~\eqref{eq:matvalHv} and the form of \( \hat{H}_z^{(+,-)}(x,y) \) are derived in Appendix~\ref{app:ValleyHamiltonian} in greater detail. Finally, Appendix~\ref{app:PhaseExtr} contains a prescription on how to determine the quantities $\phi(x,y)$ and $\Delta(x,y)$.

\section{Computational Implementation}
\label{sec:comp_imp}

\subsection{Potential Generation}
\label{sec:potential_generation}

The external electrostatic potential, which defines both lateral and vertical confinement in the device, is obtained by solving the Laplace equation with position-dependent permittivity,
\[
\nabla \cdot \left[ \epsilon(\mathbf{r}) \nabla \phi(\mathbf{r}) \right] = 0,
\]
where \( \epsilon(\mathbf{r}) \) is the local dielectric constant. This allows us to account for dielectric discontinuities between different materials in the device. In particular, we use \( \epsilon_{\text{Si}} = 11.7 \, \epsilon_0 \) for silicon and \( \epsilon_{\text{SiO}_2} = 3.9 \, \epsilon_0 \) for silicon dioxide. These values are used throughout for the charge-based (non-valley) calculations presented in this work.

The electrostatic problem is solved on a uniformly spaced three-dimensional grid using Gauss-Seidel relaxation with overrelaxation to accelerate convergence. Gate electrodes are modeled as metallic blocks held at fixed potentials (Dirichlet boundary conditions), while all other boundaries are treated with Neumann (zero normal derivative) conditions. This setup captures the essential electrostatic environment of a double quantum dot device, including oxide thickness and gate layout.

\subsection{Solutions to the stationary Schrödinger Equation}

We summarize the essential aspects of the computational framework used for both single- and two-electron quantum dot simulations. Our implementation supports both effective 2D Hamiltonians derived from the Born-Oppenheimer projection method and full 3D calculations, using FFT-based kinetic operators, matrix-free solvers, and precomputed Coulomb kernels.
For one-electron problems, the Hamiltonians used are those derived in Eqs.~\eqref{eq:effective_charge_ham} or \eqref{eq:H_eff_final}, depending on whether valley effects are included. For two-electron systems, we solve the Schr\"odinger equation for the total Hamiltonian,
\begin{equation}
\label{eq:2e-Ham}
\hat{H}_{\text{tot}} = \hat{H}_{\text{eff}}^{(1)} \otimes \mathbb{I} + \mathbb{I} \otimes \hat{H}_{\text{eff}}^{(2)} + \hat{V}_{C},
\end{equation}
with the Coulomb interaction,
\begin{equation}
V_C(\mathbf{r}_1, \mathbf{r}_2) = \frac{e^2}{4\pi\varepsilon_0\varepsilon_r |\mathbf{r}_1 - \mathbf{r}_2 + \delta|},
\end{equation}
where \( \varepsilon_r = 11.9 \) for silicon and \( \delta = 0.1 \)~nm ensures regularization for coinciding coordinates.

The real-space domain is discretized uniformly along each spatial direction, forming a finite grid of \( N_i \) points over a domain of length \( L_i \), where \( i = x,y \) (in 2D) or \( i = x,y,z \) (in 3D). The corresponding position-space grid is defined by
\[
x_i^{(n)} = n_i \frac{L_i}{N_i}, \quad n_i = 0, 1, \ldots, N_i - 1,
\]
which imposes periodic boundary conditions and ensures compatibility with discrete Fourier transform conventions.

To represent the kinetic energy operator efficiently, we work in reciprocal (momentum) space, where it becomes diagonal. The momentum-space (Fourier) grid is defined as
\[
k_i = \frac{2\pi}{L_i} \left(n_i - \frac{N_i}{2} \right), \quad n_i = 0, 1, \ldots, N_i - 1,
\]
which centers the origin of momentum space and aligns with standard FFT conventions. Since the discrete Fourier transform maps between grids of the same size, the number of grid points \( N_i \) is the same in both real and momentum space; thus, we use the same symbol to refer to both.

The kinetic energy operator, defined from the band dispersion in Eq.~\eqref{eq:eff_bandstructure}, acts in real space via the Fourier operators as
\begin{equation}
\hat T\,\psi \;=\; \big(\mathcal F^{-1}\,\epsilon\,\mathcal F\,\psi\big)(\mathbf r).
\end{equation}
where \( \mathcal{F} \) and \( \mathcal{F}^{-1} \) denote the forward and inverse Fourier transforms, respectively. These transforms are performed using FFTs, which allow for efficient and accurate switching between real and momentum space representations.
For silicon quantum dots, the kinetic term reflects the anisotropic effective masses \( m_x = m_y = m_t = 0.19 m_e \) and \( m_z = 0.98 m_e \), which define the curvature of \( \epsilon(\mathbf{k}) \) in each direction.

The full eigenvalue problem is solved using the Lanczos algorithm for Hermitian operators, as implemented in the \texttt{scipy.sparse.linalg.eigsh} routine~\cite{scipy2024}. Our implementation is fully matrix-free: the Hamiltonian is applied directly to wavefunctions without ever forming the full matrix. Each application of the Hamiltonian is composed of FFT-based evaluation of the kinetic term, pointwise multiplication for the potential term, and convolution or other real-space operations for Coulomb interactions. This approach significantly reduces memory requirements and enables scalable computation of a selected number of low-lying eigenstates to high precision (convergence tolerance \(10^{-8}\)).

\subsection{Valley Coupling Kernel Approximation}

From earlier works (see Ref.~\cite{hosseinkhani2020electromagnetic}) we can extract that the intervalley coupling value is approximately \(-0.26\). Since normalization requires \(C_0(\mathbf 0)=1\), we interpolate between these values by modeling \(C_0(\mathbf q)\) as the sum of two negative Gaussians centered at the critical coupling points \(\pm 2\mathbf k_0\):
\begin{equation}
C_0(\mathbf q)
= 1 + A_G \exp\!\left(-\frac{|\mathbf q-2\mathbf k_0|^2}{2\sigma^2}\right)
      + A_G \exp\!\left(-\frac{|\mathbf q+2\mathbf k_0|^2}{2\sigma^2}\right),
\end{equation}
with Gaussian width \(\sigma = 2.0~\mathrm{nm}^{-1}\) and amplitude \(A_G=-1.26\). This choice ensures
\begin{equation}
C_0(\pm 2\mathbf k_0) \simeq 1 + A_G = -0.26,
\end{equation}
because the cross-Gaussian contribution at \(\mathbf q=\pm 2\mathbf k_0\) is exponentially suppressed by
\(\exp\!\big[-(4k_0)^2/(2\sigma^2)\big]\ll 1\) for the parameters used, while
\begin{equation}
C_0(\mathbf 0)=1+2A_G\exp\!\big[-(2k_0)^2/(2\sigma^2)\big]\approx 1
\end{equation}
due to the negligible Gaussian tails at the origin. The resulting kernel is therefore close to unity away from the coupling points and suppressed near \(\pm 2\mathbf k_0\), providing a smooth, physically motivated modulation in momentum space that avoids the Gibbs-type artifacts of step-function models and remains efficient to evaluate.

\subsection{Computational Cost Analysis}

The dimensional reduction from 3D to 2D yields substantial computational savings in both memory and runtime. The dominant cost in all simulations arises from the evaluation of the kinetic energy operator via fast Fourier transforms (FFTs), which scale as \( \mathcal{O}(N \log N) \), where \( N =\prod_i N_i \) is the number of grid points.
Tables~\ref{tab:complexity_charge} and \ref{tab:complexity_valley} report the grid sizes, memory usage, and estimated execution times on a single CPU.
\begin{table}[t]
\centering
\caption{Computational complexity for charge-only quantum dot simulations (24 grid points in \( z \)).}
\label{tab:complexity_charge}
\begin{tabular}{lcccc}
\hline
Method & Grid Size & DOF (\(N\)) & Memory (MB) & Time (est.) (s) \\
\hline
2D BO (1e)   & $60 \times 12$        & $720$        & $0.01$ & \( 1 \) \\
3D Full (1e) & $60 \times 12 \times 24$ & $17,280$     & $0.26$ & \(10^{2}\) \\ 
2D BO (2e)   & $(60 \times 12)^2$     & $518{,}400$  & $7.91$ & \(10^{3}\) \\
3D Full (2e) & $(60 \times 12 \times 24)^2$ & $298{,}598{,}400$ & $4556$ & \(10^{7}\)
 \\
\hline
\end{tabular}
\end{table}
\begin{table}[t]
\centering
\caption{Computational complexity for valley-resolved simulations (100 grid points in \( z \)).}
\label{tab:complexity_valley}
\begin{tabular}{lcccc}
\hline
Method & Grid Size & DOF (\(N\)) & Memory (MB) & Time (est.) (s)\\
\hline
2D BO Valley (1e)   & $60 \times 12 \times 2$        & $1{,}440$      & $0.04$    & \( 1\)\\
3D Full Valley (1e) & $60 \times 12 \times 100$      & $72{,}000$     & $1.10$    & \(10^{2}\)
 \\
2D BO Valley (2e)   & $(60 \times 12 \times 2)^2$    & $8{,}294{,}400$ & $127$  & \( 10^3\) \\
3D Full Valley (2e) & $(60 \times 12 \times 100)^2$  & $5.18 \times 10^9$ & $79102$ & infeasible \\
\hline
\end{tabular}
\end{table}
%
%
For single-electron simulations, all 2D and 3D cases are computationally feasible. However, for two-electron systems, full 3D calculations rapidly become prohibitive. In the valley-resolved case, the memory requirement exceeds 79 GB and the runtime is beyond practical limits—rendering the 3D simulation infeasible.
In contrast, all 2D projection-based methods complete in minutes or less and remain within modest memory constraints. These models can capture charge and valley physics efficiently and accurately, making them well-equipped for large-scale simulations, device optimization, and parameter sweeps.

\section{Results}

To validate the accuracy and physical reliability of the proposed 2D projection methods, we benchmark them against full 3D quantum mechanical simulations across a range of quantum dot scenarios. As a point of comparison, we also include results from naive 2D approaches that rely on slicing the 3D potential at fixed vertical positions—methods known to fail in the presence of nontrivial vertical confinement variations.
For charge-only physics, we focus on observables that are sensitive to the precise ground state energy, such as tunnel couplings in the single-electron regime and exchange (J) couplings in the two-electron case. These quantities serve as test cases of interest to the community for the effectiveness of dimensional reduction. We perform these calculations using a realistic device potential (see Sec.~\ref{sec:dev_model}).
In the valley-resolved setting, we benchmark the projection methods by comparing computed valley splittings and valley phases against 3D references. Since valley-phase extraction requires accurate wavefunctions in addition to energies, this test directly probes the fidelity of the projected spinor structure. Here we use an artificial potential as discussed in Sec.~\ref{sec:valley_results}.
Together, these comparisons assess the capability of the 2D projection framework to accurately capture both energetic and wavefunction-level features relevant for quantum dot qubits.

\subsection{Charge-only calculations}

\subsubsection{Device Model}
\label{sec:dev_model}

\begin{figure}[t]
    \centering
    \includegraphics[width=0.75\textwidth]{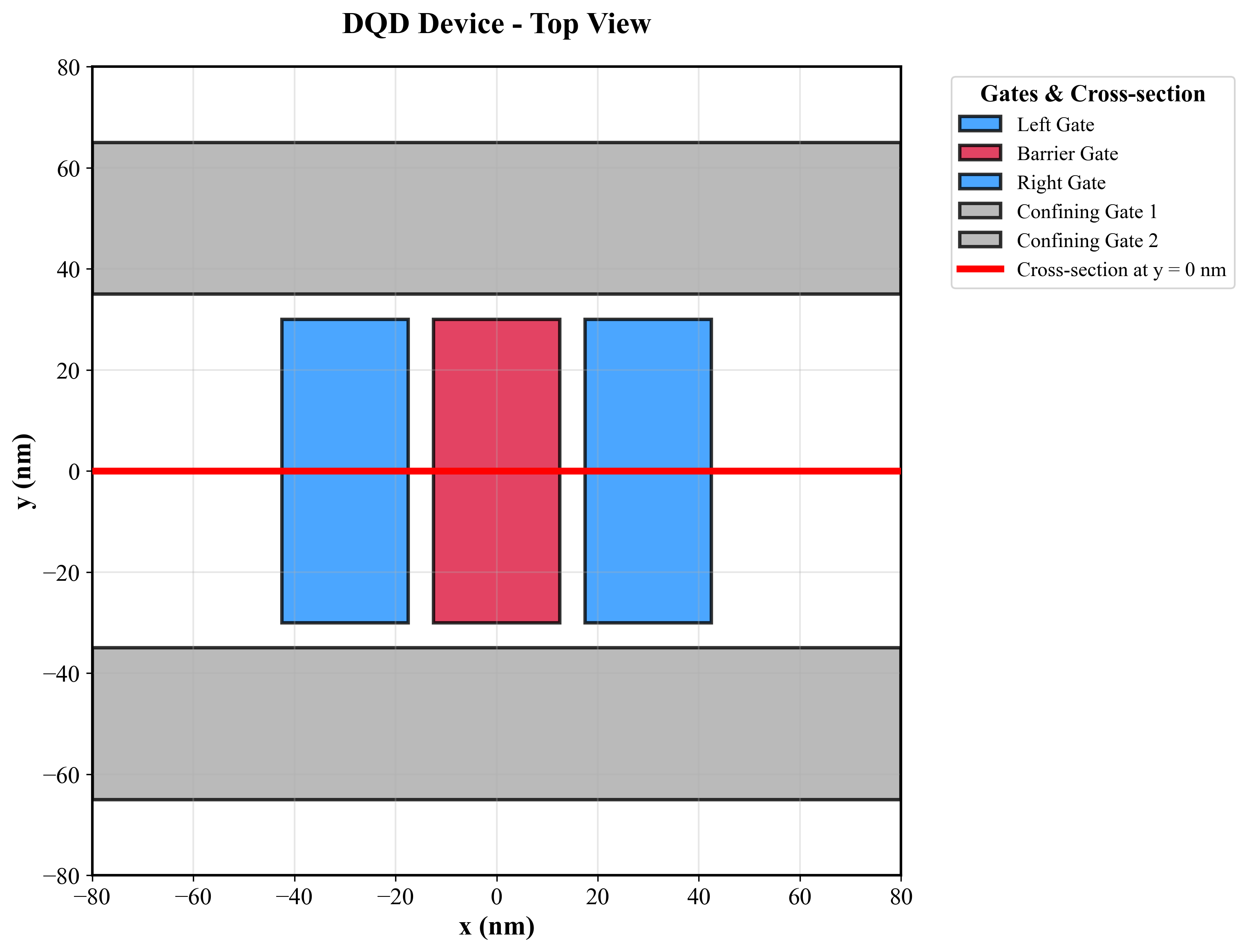}
    \caption{Top-down view of the stepped heterostructure double quantum dot. The image shows the three interface sections with different heights and the five gate electrodes. The red line at $y = 0$ indicates the position of the cross-section shown in Fig.~\ref{fig:device_cross_section}.}
    \label{fig:device_top_view}
\end{figure}

\begin{figure}[htbp]
\centering
\subfigure[Flat interface]{
    \includegraphics[width=0.47\textwidth]{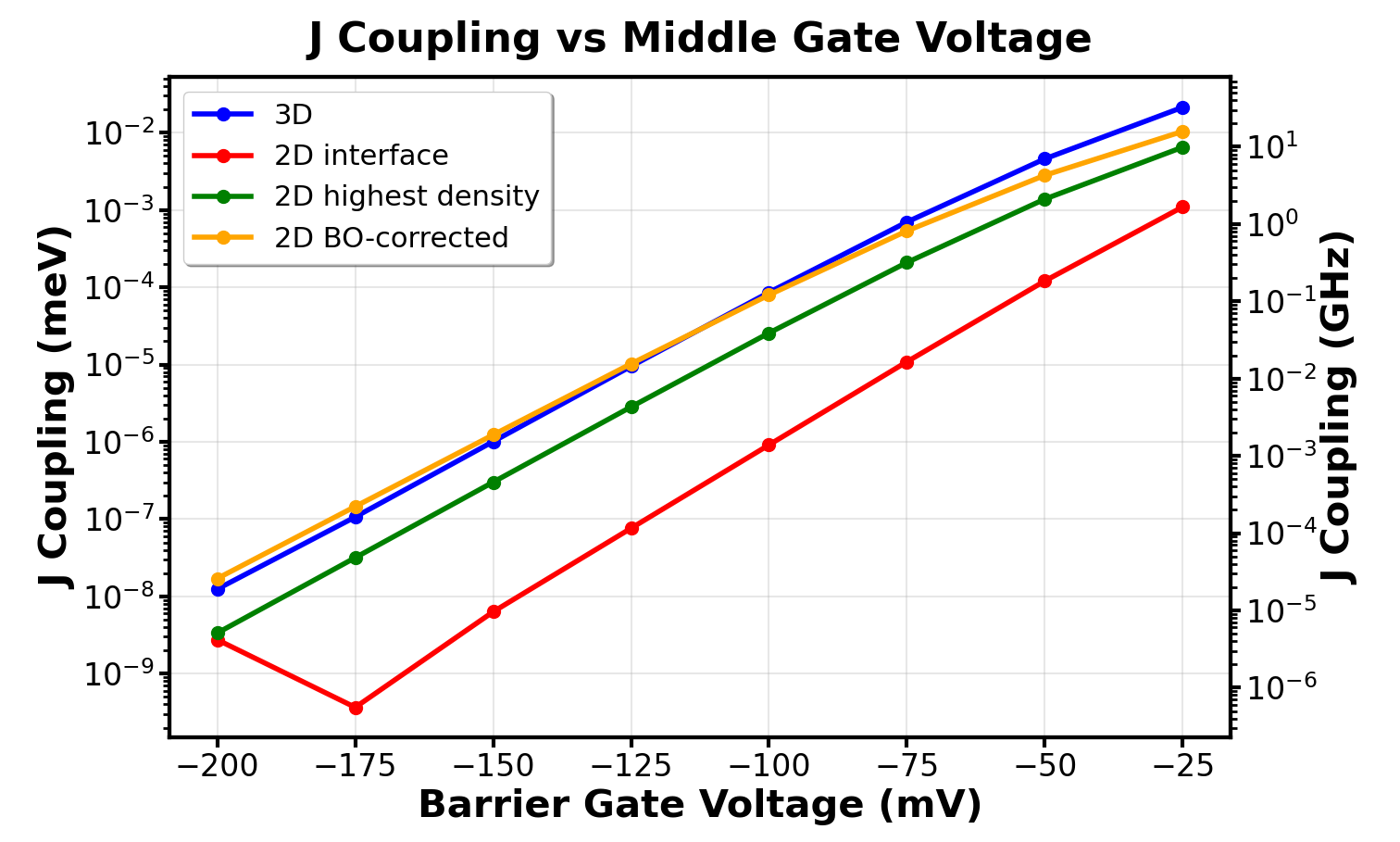}
    \label{fig:tunnel_flat}
}
\hfill
\subfigure[Interface with a step]{
    \includegraphics[width=0.47\textwidth]{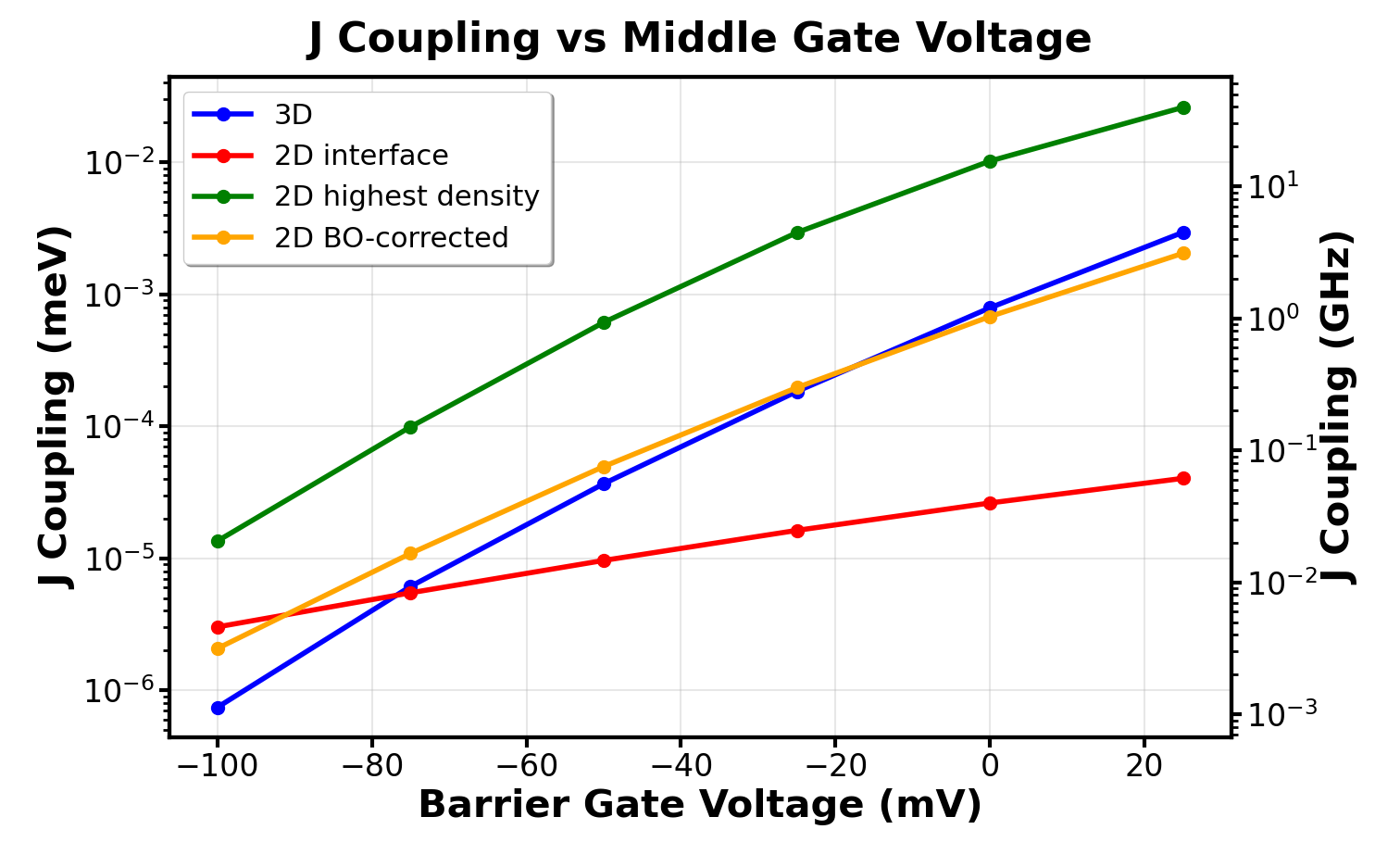}
    \label{fig:tunnel_step}
}
\caption{\( J \)-coupling versus barrier gate voltage for different dimensional approximations in double quantum dot systems. Results show: 2D slice at highest electron density (green), 2D slice at heterointerface (red), Born–Oppenheimer corrected 2D (yellow), and full 3D calculation (blue). The exponential dependence on gate voltage reflects modulation of the tunnel barrier.}
\label{fig:j_coupling_comparison}
\end{figure}

We model a double quantum dot (DQD) device consisting of metallic gates positioned above a silicon heterostructure. The device geometry is defined within a computational domain of $160 \times 160 \times 50$ nm$^3$ (see Figure \ref{fig:tunnel_step}). 
The gate architecture comprises five distinct electrodes: three plunger gates and two confinement gates, as illustrated in Fig.~\ref{fig:device_top_view}. The left and right plunger gates (P1 and P3) are positioned at $x = -30$ nm and $x = 30$ nm, respectively, each with dimensions of $25 \times 60 \times 10$ nm$^3$. The central barrier gate (P2) is located at $x = 0$ nm with dimensions $25 \times 60 \times 10$ nm$^3$ and is positioned 5 nm closer to the substrate surface than the other two gates to enhance electrostatic control over the interdot tunnel coupling. Two confining gates (C1 and C2) provide lateral confinement, positioned at $y = \pm 50$ nm with extended dimensions of $160 \times 30 \times 10$ nm$^3$.
To obtain the potential for a given gate voltage arrangement, we solve the Laplace equation as outlined in Sec.~\ref{sec:comp_imp}. We solve the latter on a grid with a resolution of 1 nm in the $x$ and $y$ directions and 0.5 nm in the $z$ direction. 

For charge-only calculations, we investigate two device configurations: a flat interface structure and a stepped heterostructure. The stepped configuration features three distinct interface sections with heights varying between $-0.5$ nm and $+0.5$ nm, representing realistic fabrication variations that arise during epitaxial growth or etching processes. In particular, variations at this magnitude should be expected as these have been observed in tunnling electron microscopy measurements of real devices, see \cite{cifuentes2024bounds}.
For this section, we assume that the material stack consists of SiO$_2$ ($\epsilon_r = 3.9$) above the interface and a silicon ($\epsilon_r = 11.9$) substrate. These varying permittivities are considered in the solution of the Laplace equation.
Additionally, there is a conduction band offset of 3 eV  between the Si and SiO$_2$ regions. We add this value to the gate-defined electrostatic potential in the SiO$_2$ domain (all space above $z > 0$) to confine the quantum dot in the $z$-direction in a realistic manner.

\begin{figure}[t]
    \centering
    \includegraphics[width=0.85\textwidth]{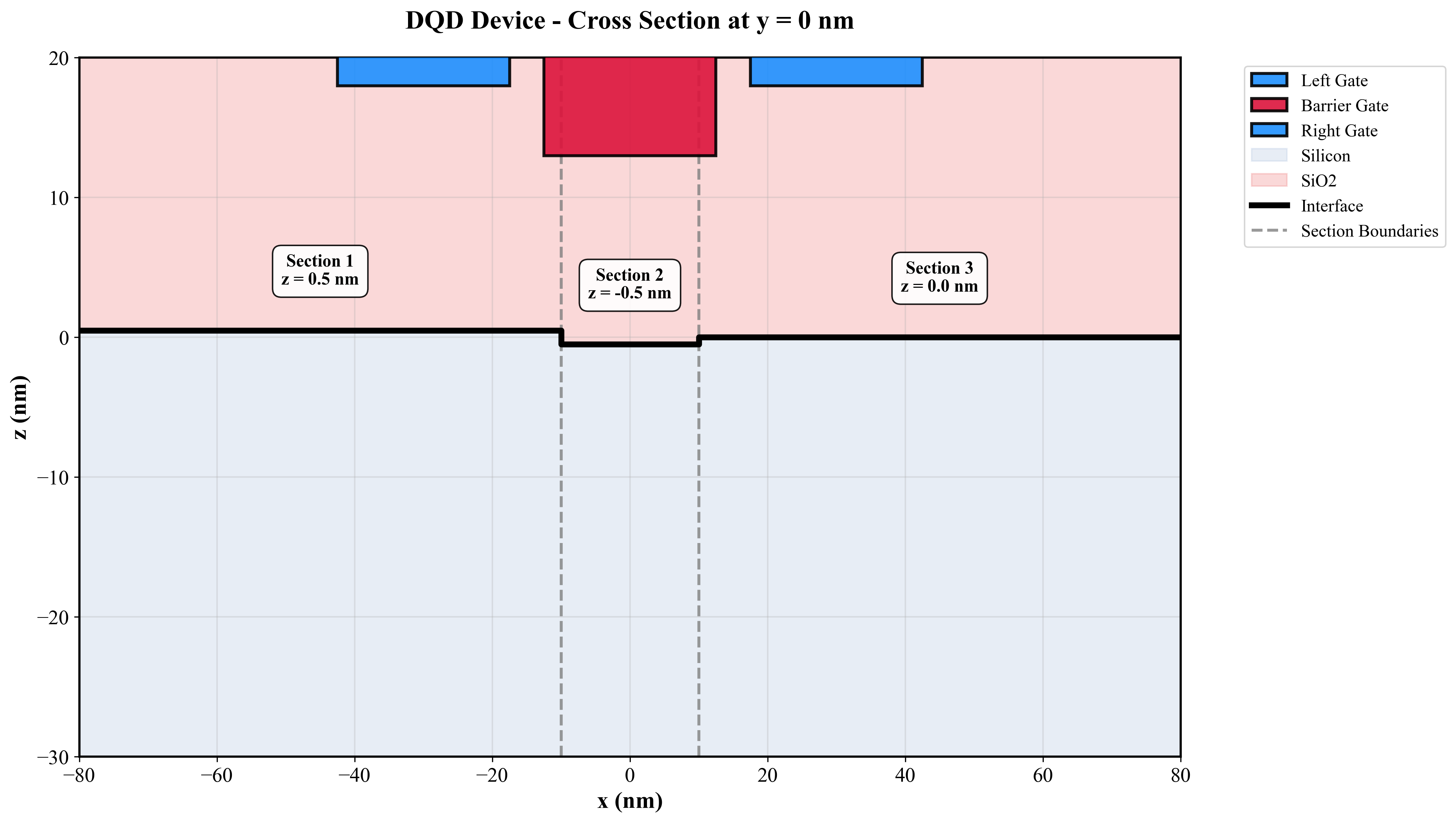}
    \caption{Cross-sectional view of the stepped heterostructure double quantum dot along the $xz$-plane at $y = 0$. The stepped interface lies beneath the three main gates (left plunger, barrier, and right plunger). It consists of three sections: Section 1 ($x \in [-80, -10]$ nm) at $z = 0.5$ nm, Section 2 ($x \in [-10, +10]$ nm) at $z = -0.5$ nm, and Section 3 ($x \in [+10, +80]$ nm) at $z = 0$ nm. Silicon lies below the interface while SiO$_2$ is above.}
    \label{fig:device_cross_section}
\end{figure}


\subsubsection{Tunnel Coupling Analysis}
\label{sec:tunnel_coupling}

We systematically investigate tunnel couplings in double quantum dot systems using three different computational approaches. The tunnel coupling represents the energy scale for coherent charge transfer between quantum dots and is computed as half the energy difference between the first two eigenstates in charge space: $t = |E_1 - E_0|/2$, where -- in our case -- $E_0$ and $E_1$ are the ground and first excited state eigenvalues, obtained from the single-particle Hamiltonian $H = T + V(\mathbf{r})$, e.g., using Lanczos eigensolvers. 
The double quantum dot system is modeled a 120~nm $\times$ 24~nm domain in the $xy$-plane with vertical extent from -10~nm to +2~nm, discretized on a 60$\times$10$\times$24 grid for 3D calculations. The five-gate configuration consists of left and right plunger gates (initially 400~mV), a central barrier gate (scanned from -120 to 60~mV), and top/bottom gates (-1000~mV) providing vertical confinement.
We compare four computational approaches for evaluating single-particle properties:
\begin{enumerate}
    \item \textbf{Full 3D calculation:} Solves the full single-particle Schrödinger equation using the three-dimensional Hamiltonian~\eqref{eq:charge_ham}.
    
    \item \textbf{2D slice at the interface:} Uses the effective 2D Hamiltonian~\eqref{eq:effective_charge_ham}, with the potential taken from a fixed slice at the heterointerface (\(z = -0.5~\text{nm}\)).
    
    \item \textbf{2D slice at maximum density:} Also uses Eq.~\eqref{eq:effective_charge_ham}, but with the potential taken from the position of maximum electron density (\(z = -2.5~\text{nm}\)), and neglects the confinement correction term \(\epsilon_0(x, y)\).
    
    \item \textbf{Born–Oppenheimer-corrected 2D calculation:} Uses the same effective Hamiltonian~\eqref{eq:effective_charge_ham}, but includes the confinement correction \(\epsilon_0(x, y)\), obtained by solving the transverse Schrödinger equation at each lateral position.
\end{enumerate}

For each barrier gate voltage, the right plunger gate is optimized within a 200-500~mV range to minimize the energy gap $\Delta E$ between ground and excited state using Nelder-Mead optimization \cite{nelder1965simplex}. We then identify $\frac{\Delta E}{2}$ as the tunnel coupling $t$.
\begin{figure}[htbp]
    \centering
    \subfigure[Flat interface]{
        \includegraphics[width=0.47\textwidth]{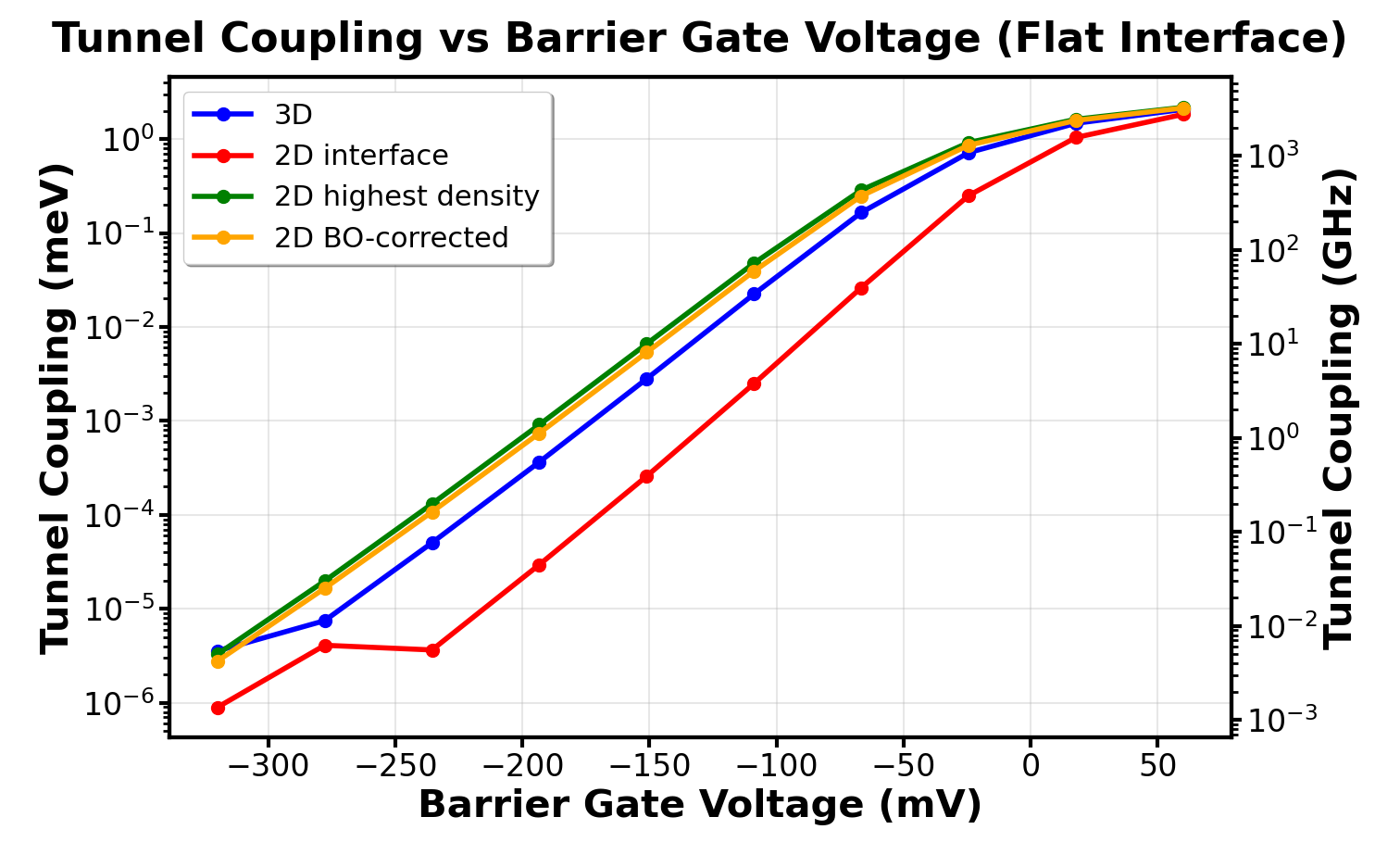}
        \label{fig:tunnel_coupling_flat}
    }
    \hfill
    \subfigure[Stepped interface]{
        \includegraphics[width=0.47\textwidth]{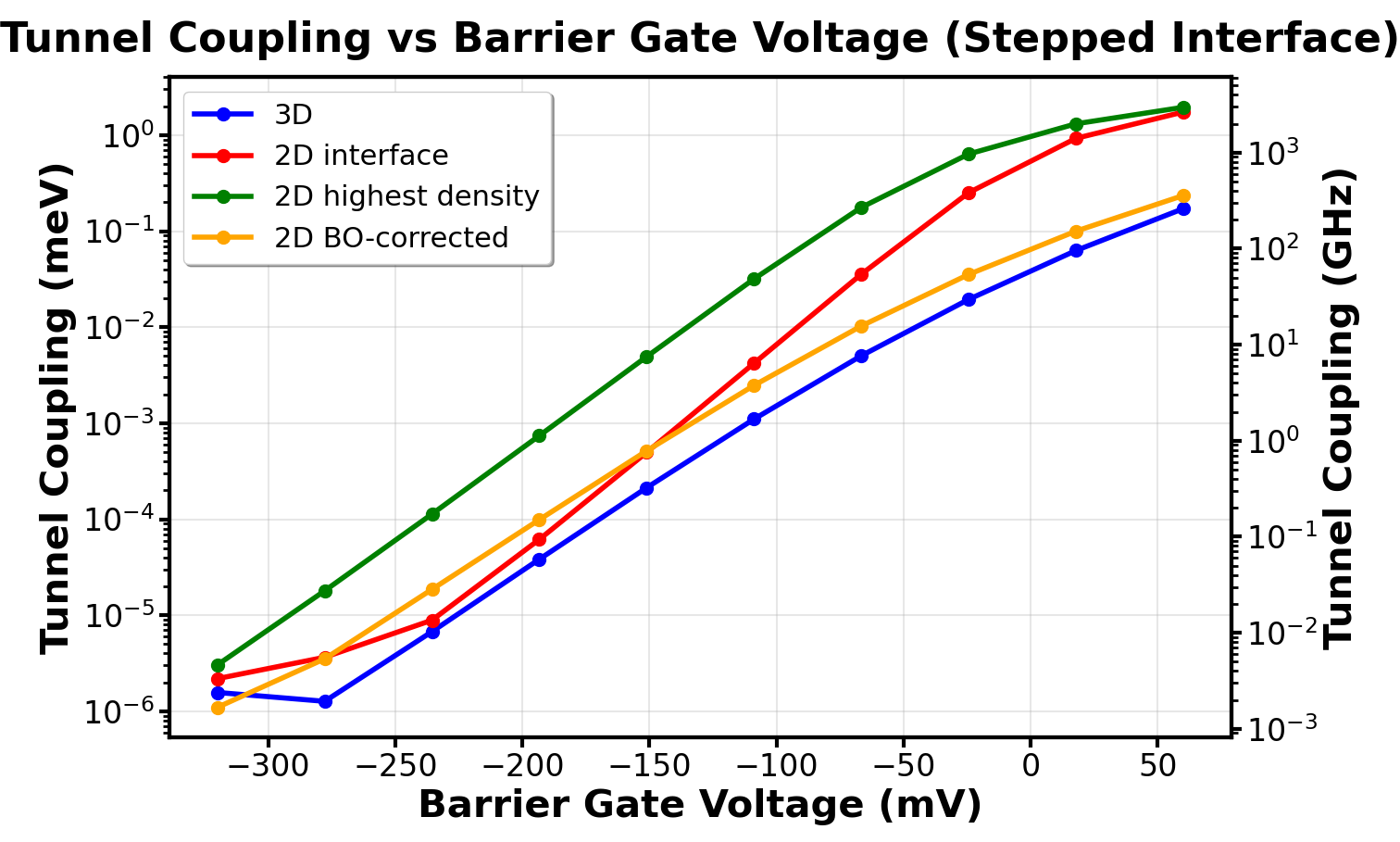}
        \label{fig:tunnel_coupling_step}
    }
    \caption{Tunnel coupling vs.\ barrier gate voltage for flat and stepped heterointerfaces. Methods: 3D (blue dots), 2D interface slice (red dots), highest density slice (green dots), and Born-Oppenheimer corrected 2D (orange dots). The Born-Oppenheimer correction captures vertical confinement with 2D efficiency, matching full 3D results.}
    \label{fig:tunnel_coupling_scan}
\end{figure}
For the flat interface case, the 2D slice at the interface position deviates by approximately one order of magnitude from the benchmark 3D calculation. In contrast, the 2D slice at the highest density position shows deviations of at most a factor of two from the correct result, comparable to the Born-Oppenheimer (BO) treatment,i which exhibits slightly better agreement with the benchmark. This suggests that the effects of varying vertical confinement across the device are relatively mild for our given device, making a straightforward 2D slicing approach sufficient for reasonable accuracy.

However, for the stepped interface case, the highest density approach performs poorly, with deviations up to two orders of magnitude from the correct 3D result. The interface slice method shows marginal improvement but still deviates by more than one order of magnitude in several cases. In contrast, the projection treatment remains consistently close to the benchmark result, with deviations typically limited to a factor of two or less. This demonstrates that the projection method effectively corrects for interface-induced effects that severely impact conventional 2D approaches.
These findings indicate that while simple 2D slicing may suffice for relatively smooth interfaces, the projection correction becomes essential for accurately capturing vertical confinement effects in the presence of interface roughness. The projection method thus provides a computationally efficient 2D approach that maintains the accuracy of full 3D calculations, even for challenging interface geometries.

\subsubsection{Exchange (J) Coupling Calculations}

Next, we systematically compare \( J \)-coupling calculations again using four approaches. The double quantum dot system spans a \( 120~\text{nm} \times 12~\text{nm} \) lateral domain with vertical extent from \(-10~\text{nm}\) to \(+2~\text{nm}\), discretized on a \(30 \times 3 \times 24\) grid for 3D calculations.
The \( J \)-coupling represents the exchange interaction energy between electrons and is computed as the energy splitting between the ground and first excited states of a two-electron system: \( J = |E_1 - E_0| \). We obtain \( E_0 \) and \( E_1 \) as the two lowest eigenvalues of the two-electron Hamiltonian \eqref{eq:2e-Ham}. For the 2D case, we use the projected effective potential defined in Eq.~\eqref{eq:effective_charge_ham} and a two-dimensional Coulomb kernel. Further computational details are provided in Sec.~\ref{sec:comp_imp}.

The four variants of determining the \( J \)-coupling used here are - apart from using \eqref{eq:2e-Ham} instead of \eqref{eq:effective_charge_ham} - identical to those presented in Sec.~\ref{sec:tunnel_coupling}. In particular, \( J \)-couplings are obtained from a scan of the barrier gate voltage from \(-200~\text{mV}\) to \(-25~\text{mV}\), while keeping the plunger gates fixed at \(400~\text{mV}\) and the screening gates at \(-1000~\text{mV}\).

In the flat-interface case, the 2D slice at the heterointerface deviates by about one order of magnitude or more from the 3D benchmark. The 2D slice at the highest density performs better, with deviations typically within a factor of two. In contrast, the Born–Oppenheimer-corrected potential yields excellent agreement across the entire scan range, with deviations generally under 50\%.

An even clearer picture emerges for the stepped interface. The corrected potential method closely matches the 3D results, while both slicing methods exhibit significant errors—typically one to two orders of magnitude. Notably, the heterointerface slice fails to capture the correct trend altogether.

\subsection{Valley resolved calculations}
\label{sec:valley_results}

In this section, we investigate static valley-state properties as a function of the quantum dot position as it is moved across atomically sharp interfaces and smooth interfaces generated by germanium doping. The two metrics we choose for this benchmarking task are valley splitting and valley phases as defined in \cite{langrock2023blueprint}.
We calculate these quantities within our full 3$D$ description \eqref{eq:env_function_theory} and compare the results to the approximate 2$D$ methods as described in Eq.~\eqref{app:ValleyHamiltonian}.
In this section, we use an artificial potential, combining the harmonic lateral confinement, linear vertical confinement, and an additional potential that depends on the choice of materials,
\begin{equation}
\label{eq:qdot_potential}
V(x,y,z) = \frac{1}{2}[\omega_x^2(x-x_0)^2 + \omega_y^2 y^2] - \alpha z + V_{\text{add}}(x,y,z),
\end{equation}
where we choose $\omega_x = \omega_y = 1.0$~meV/nm$^2$ and $\alpha = 5$~meV/nm, corresponding to a quantum dot size of about $10$ to $15$~nm. $V_{\text{add}}(x,y,z)$ labels the additional potential from material variations, which is responsible for confining the electron to the silicon channel. 
For both interface choices, we perform a scan over $x_0$ -- and hence the quantum dot position -- allowing systematic characterization of valley phase and splitting variations.

\subsubsection{Atomistic steps}

The additional potential \( V_{\text{add}} \) in this framework is chosen as
\begin{equation}
V_{\text{add}}(x,y,z) = 3000\,\text{meV} \times \Theta\big(z - h_{\text{step}}(x,y)\big),
\end{equation}
where \(\Theta\) is the Heaviside step function and the interface position function \(h_{\text{step}}(x,y)\) is defined as
\[
h_{\text{step}}(x,y) = 
\begin{cases}
0.135\,\text{nm}, & x \in [-4.7, 5.7]\,\text{nm}, \\
0, & \text{elsewhere}.
\end{cases}
\]
In practice, we approximate the Heaviside function \(\Theta\) with a smooth sigmoid function to model the interface transition with atomic-scale smoothness. Specifically, we use the sigmoid function,
\begin{equation}
S(z; z_0, w) = \frac{1}{1 + \exp\left[-4\frac{z - z_0}{w}\right]},
\end{equation}
where \(z_0 = h_{\text{step}}(x,y)\) defines the interface position and \(w = 0.135\,\text{nm}\), the transition width, is chosen to match the length scale of adjacent horizontal atomic layers in silicon. This smooth approximation eliminates discontinuities, improving numerical stability and convergence, while capturing the sub-nanometer scale interface broadening observed experimentally.

\begin{figure}[htbp]
    \centering
    \includegraphics[width=0.7\textwidth]{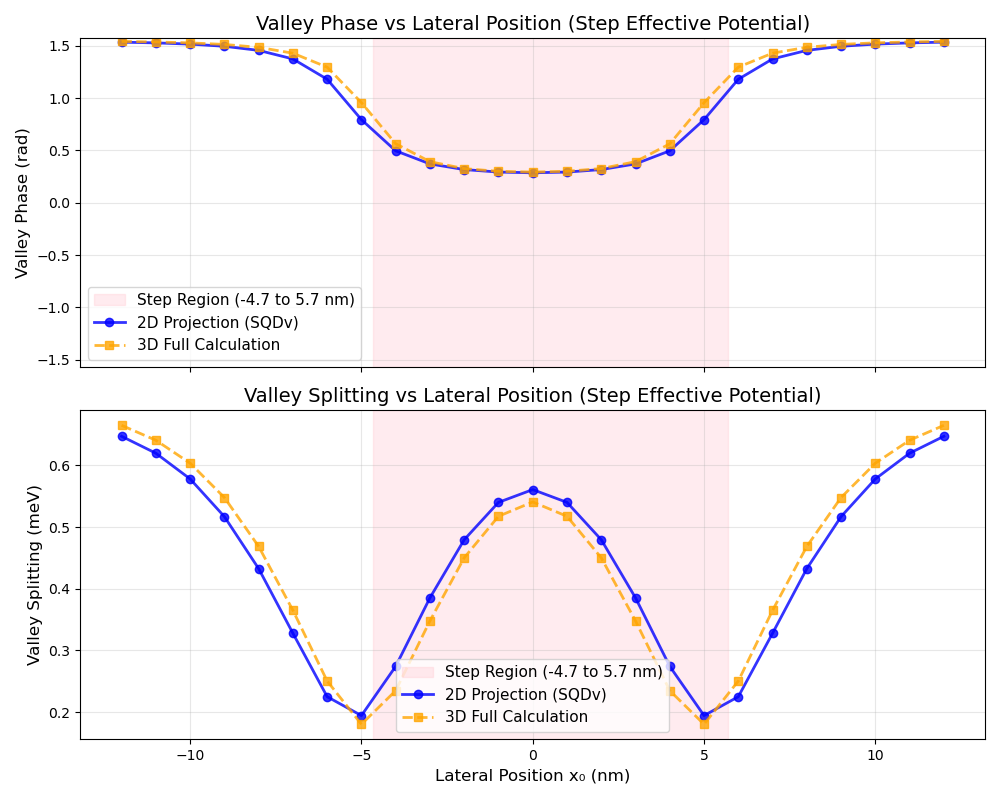}
    \caption{Valley phase and valley splitting across an atomically sharp interface step. (Upper plot) Valley phase and (lower plot) valley splitting versus lateral quantum dot position $x_0$. Blue circles show 2D projection results, orange dots show full 3D calculations. The pink shaded region indicates the step location. All methods capture the systematic valley phase modulation with excellent quantitative agreement.}
    \label{fig:valley_scan_step}
\end{figure}

Figure~\ref{fig:valley_scan_step} demonstrates excellent agreement between valley splittings obtained with the projection method \eqref{eq:matvalHv} and the full 3D treatment using the Hamiltonian \eqref{eq:gef_schrodinger}. As expected, the valley phase exhibits a systematic modulation across the step region with a characteristic amplitude of approximately \(\pi/3\) radians, while the valley splittings vary complementarily from 0.15 to 0.7 meV. When the electron is not close to the step, the valley splitting values are comparable to those reported in the literature for flat surfaces \cite{Boykin2004}.

\subsubsection{Smooth interfaces (SiGe)}

\begin{figure}[h!]
    \centering
    \includegraphics[width=0.6\textwidth]{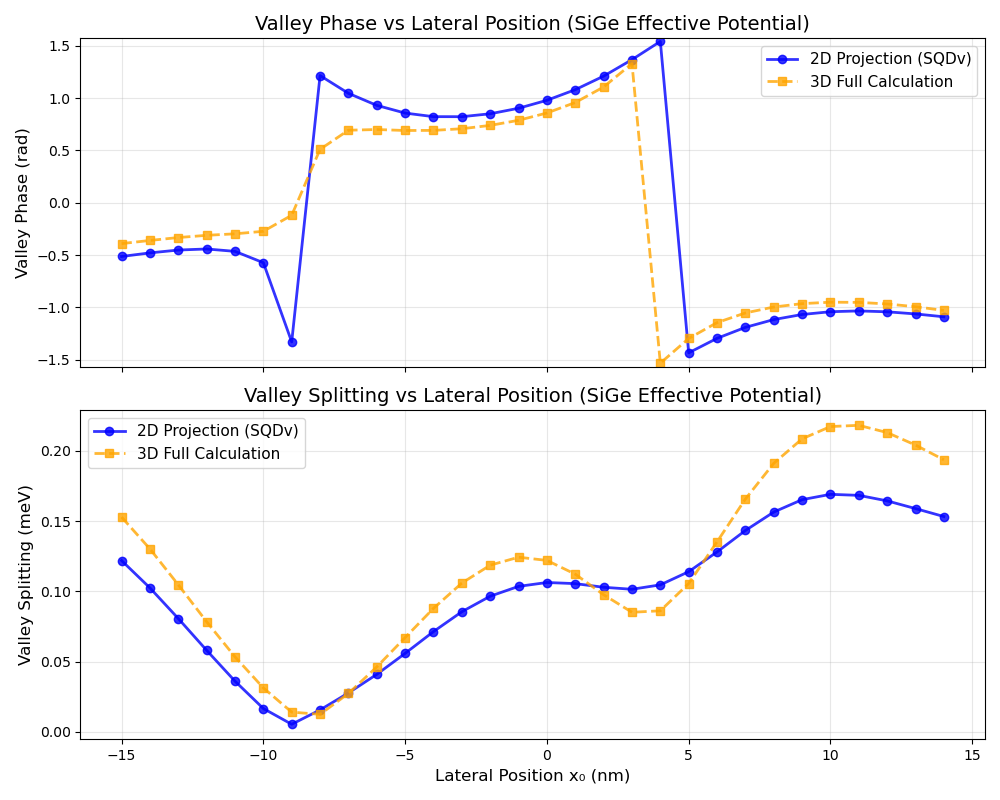}
    \caption{Comparison of valley phase (top) and valley splitting (bottom) between the 2D projection method (blue dots) and full 3D calculations (orange squares) across lateral dot positions \( x_0 \).}
    \label{fig:valley_sige_results}
\end{figure}

We now turn to a different type of hetero-interface, formed between a Si quantum well and a SiGe barrier.  Interdiffusion of Ge atoms at the barrier lead to a gradual change of the Ge concentration across the interface region.
We introduce the atomistic disorder through random sampling of Ge concentration \cite{lima2023interface}. The local SiGe potential which serves as the additional potential in \eqref{eq:qdot_potential} is computed from the sampled Ge content via \( V_{\text{SiGe}}(x,y,z) = V_{\text{max}} \cdot \frac{N_{\text{Ge}}(x,y,z)}{N_{\text{total}} \cdot c_{\text{max}}} \), where \( V_{\text{max}} = 150 \) meV sets the strength of the alloy potential for a Ge content of x=0.3. 
\begin{figure}[htbp]
    \centering
    \includegraphics[width=0.85\textwidth]{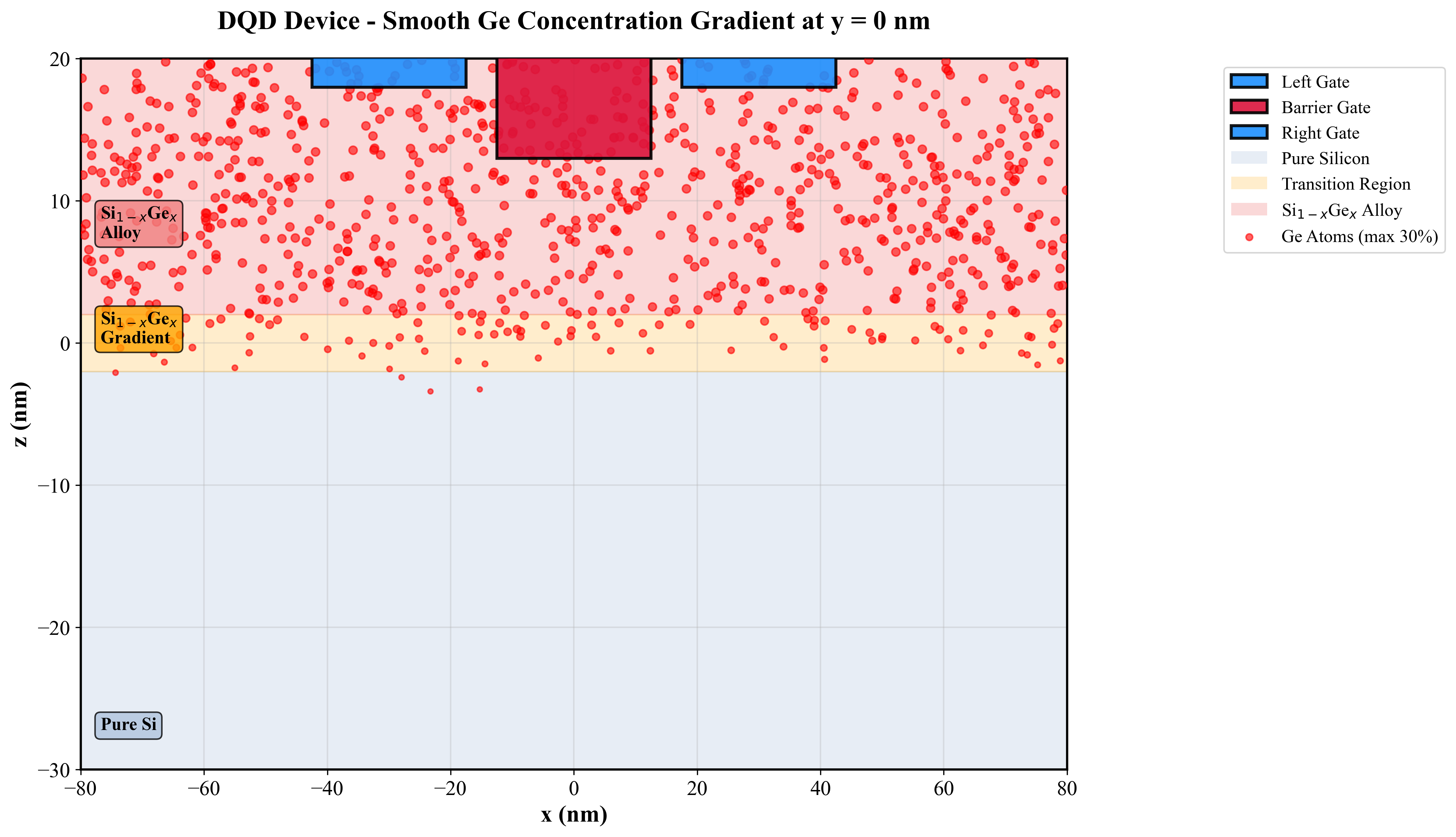}
    \caption{Cross-sectional view of the stepped heterostructure with smooth Ge concentration gradient. The plot shows the device geometry in the $xz$-plane at $y = 0$ nm (see Fig.~\ref{fig:device_top_view}), displaying the three main gate electrodes (left plunger, barrier, and right plunger gates) positioned above a realistic Si$_{1-x}$Ge$_x$ heterostructure. Red dots represent individual Ge atoms with local concentration following a smooth sigmoid profile that transitions from pure Silicon (below $z = -2$ nm) to Si$_{0.7}$Ge$_{0.3}$ alloy (above $z = +2$ nm). }
\label{fig:sige_heterostructure}
\end{figure}
Figure~\ref{fig:valley_sige_results} shows the computed valley phase and valley splitting as a function of lateral dot position \( x_0 \). The 2D method reproduces the 3D results with high fidelity in splitting and captures key features in the valley phase, despite its reduced dimensionality.
The dot center \( x_0 \) is varied from \(-11\) to \(+8\) nm in 0.25 nm steps, producing 77 sampling positions. Simulations are performed on a domain spanning \( x \in [-15,15] \) nm, \( y \in [-10,10] \) nm, and \( z \in [-10.75, 2.75] \) nm, discretized on a \( 30 \times 10 \times 200 \) grid. A random seed ensures reproducibility of the alloy disorder.
A detailed description of the alloy model, sampling procedure, potential generation, and numerical implementation is provided in Appendix~\ref{apx:sige_valley}.
The 2D projection method again achieves good agreement in valley splitting and captures the key spatial variation in the valley phase. Although some fine-grained phase structure is smoothed out compared to 3D, the overall trend and correlation are preserved. Given its substantial computational speedup, our effective 2D-multivalley envelope function theory provides a powerful tool for valley landscape modeling in disordered SiGe systems, comparted to 3D approaches. 

\newpage
\section{Discussion and Outlook}

The accurate and efficient simulation of silicon spin qubits relies on a productive feedback loop between theory (simulation) and experiment. As the field matures, the ability to iterate quickly between device proposals, modeling, fabrication, and measurement becomes increasingly important. In this context, computational efficiency becomes a critical bottleneck. We have shown that charge-only observables such as tunnel and exchange couplings are reproduced with high accuracy using the projected 2D models, even in the presence of vertical confinement variations and interface roughness. For valley-resolved calculations, the method allows efficient and reliable extraction of the valley splitting and phase in both sharp and disordered interfaces, with accuracy comparable to full 3D simulations. Further work will be required to apply the presented computational framework to fully disordered, rough Si/SiO interfaces in Si-MOS quantum dot devices.

Crucially, two-electron valley simulations, previously infeasible, are now within reach using modest computational resources. This opens the door to studying spin-valley dynamics in realistic device geometries, modeling gate operations, and assessing the impact of valley physics on qubit performance. With accurate two-electron valley simulations now tractable, the framework makes previously inaccessible regimes computationally feasible. It further lays the groundwork for modeling spin-related properties such as positionally dependent $g$-factors in realistic geometries. This will require future work, incorporating the spin DOF and spin-orbit coupling into the presented framework. Moreover, the substantial efficiency gains make the approach well-suited for time-dependent simulations, e.g., describing qubit shuttling \cite{mills2019shuttling,noiri2022shuttling,kunne2024spinbus,de2025high,jeon2025, bindershuttling}, and may even enable exploration of few-electron systems beyond the two-electron limit.

In summary, the approach presented here enables efficient, accurate, and physically grounded simulations of silicon quantum dots, supporting faster theory–experiment feedback cycles and advancing the design of next-generation qubit devices.

\newpage

\appendix
\section{Derivation of the defining equation for the generalized envelope function}
\label{app:app_der_mpf}

In the following, we derive the effective Schrödinger equation satisfied by the \textit{generalized envelope function} $\psi(\mathbf{r})$, introduced in the main text. 

We begin with the single-particle Schr\"odinger equation in the Bloch basis
\begin{equation}
\sum_{\mathbf k}\!\big[V(\mathbf r)+\epsilon_{\mathbf k}\big]\,
\phi_{\mathbf k}(\mathbf r)\,c_{\mathbf k}
\;=\;
E \sum_{\mathbf k}\phi_{\mathbf k}(\mathbf r)\,c_{\mathbf k},
\end{equation}
with
\begin{equation}
\Psi(\mathbf r)=\sum_{\mathbf k}c_{\mathbf k}\,\phi_{\mathbf k}(\mathbf r),
\qquad
\phi_{\mathbf k}(\mathbf r)=u_{\mathbf k}(\mathbf r)\,e^{i\mathbf k\cdot\mathbf r}.
\end{equation}
Projecting onto $u_{\mathbf k'}^*(\mathbf r)e^{-i\mathbf k'\cdot\mathbf r}$ over a crystal volume $\Omega$ yields
\begin{equation}
\sum_{\mathbf k}\int_{\Omega}\! d^3\mathbf r\;
u_{\mathbf k'}^*(\mathbf r)\,u_{\mathbf k}(\mathbf r)\,V(\mathbf r)\,e^{i(\mathbf k-\mathbf k')\cdot\mathbf r}\,c_{\mathbf k}
\;+\;\delta_{\mathbf k,\mathbf k'}\,\epsilon_{\mathbf k}\,c_{\mathbf k}
\;=\;E\,c_{\mathbf k'}.
\label{eq:dis_schr}
\end{equation}

Expanding $u_{\mathbf k}(\mathbf r)=\Omega^{-1/2}\sum_{\mathbf G} c_{\mathbf G}^{(\mathbf k)}\,e^{i\mathbf G\cdot\mathbf r}$ and keeping the $\mathbf G=\mathbf G'$ terms (as in Ref.~\cite{hosseinkhani2020electromagnetic}) gives the approximation
\begin{equation}
\frac{1}{\Omega}\int d^3\mathbf r\;
u_{\mathbf k'}^*(\mathbf r)\,u_{\mathbf k}(\mathbf r)\,V(\mathbf r)\,e^{i(\mathbf k-\mathbf k')\cdot\mathbf r}
\;\approx\; C_{0,\mathbf k,\mathbf k'}\,\widetilde V_{\mathbf k-\mathbf k'}. 
\end{equation}
If we assume that terms with $\mathbf{G} = \mathbf{G}'$ dominate (as done in earlier works \cite{hosseinkhani2020electromagnetic}), we may define the approximate overlap kernel
\begin{equation}
C_{0,\mathbf k,\mathbf k'}=\frac{1}{\Omega}\sum_{\mathbf G}
c_{\mathbf G}^{(\mathbf k)}\,c_{\mathbf G}^{(\mathbf k')*}.
\end{equation}
At this point we choose the difference-only approximation
\begin{equation}
C_{0,\mathbf k,\mathbf k'}\;\approx\;C_{0,\mathbf k-\mathbf k'},
\end{equation}
i.e., we assume the dominant overlap is translationally invariant in $\mathbf k$-space. Using this, the Schrodinger equation in k-space \eqref{eq:dis_schr} becomes
\begin{equation}
\sum_{\mathbf k} C_{0,\mathbf k-\mathbf k'}\,\widetilde V_{\mathbf k-\mathbf k'}\,c_{\mathbf k}
\;+\;\delta_{\mathbf k,\mathbf k'}\,\epsilon_{\mathbf k}\,c_{\mathbf k}
\;=\;E\,c_{\mathbf k'}.
\end{equation}

Passing to the continuum one obtains
\begin{equation}
\int \frac{d^3\mathbf k}{(2\pi)^{3/2}}\,
C_{0}(\mathbf k-\mathbf k')\,\widetilde V(\mathbf k-\mathbf k')\,c(\mathbf k)
\;+\;\epsilon(\mathbf k')\,c(\mathbf k')\;=\;E\,c(\mathbf k').
\end{equation}

We now define the Fourier operators (symmetric normalization) once, compactly, and use them throughout:
\begin{equation}
(\mathcal F\psi)(\mathbf k)=\int \frac{d^3\mathbf r}{(2\pi)^{3/2}}\,e^{-i\mathbf k\cdot\mathbf r}\,\psi(\mathbf r),
\qquad
(\mathcal F^{-1}c)(\mathbf r)=\int \frac{d^3\mathbf k}{(2\pi)^{3/2}}\,e^{+i\mathbf k\cdot\mathbf r}\,c(\mathbf k),
\qquad
\psi=\mathcal F^{-1}c,\quad c=\mathcal F\psi.
\end{equation}

We insert identities $\mathcal F\mathcal F^{-1}$ and apply $\mathcal F^{-1}$ to the $k$-space equation to get the operator form
\begin{equation}
\mathcal F^{-1}C_0\widetilde V\,\mathcal F\,\psi \;+\; \mathcal F^{-1}\epsilon\,\mathcal F\,\psi \;=\; E\,\psi.
\end{equation}
To make the potential term explicit, we keep the $\mathbf k'$ integral and introduce $\mathbf q=\mathbf k-\mathbf k'$:
\begin{align}
\big[\mathcal F^{-1}C_0\widetilde V\,\mathcal F\,\psi\big](\mathbf r)
&=\int d^3\mathbf r'\,\psi(\mathbf r')\!
\int \frac{d^3\mathbf k\,d^3\mathbf k'}{(2\pi)^{9/2}}\,
C_0(\mathbf k-\mathbf k')\,\widetilde V(\mathbf k-\mathbf k')\,e^{i\mathbf k'\cdot\mathbf r-i\mathbf k\cdot\mathbf r'} \nonumber\\
&=\int d^3\mathbf r'\,\psi(\mathbf r')\!
\int \frac{d^3\mathbf q\,d^3\mathbf k'}{(2\pi)^{9/2}}
\,C_0(\mathbf q)\widetilde V(\mathbf q)\,e^{i\mathbf k'\cdot(\mathbf r-\mathbf r')}e^{-i\mathbf q\cdot\mathbf r'}
\qquad(\mathbf k=\mathbf k'+\mathbf q)\nonumber\\
&=\int d^3\mathbf r'\,\psi(\mathbf r')\,\underbrace{\left[\int \frac{d^3\mathbf k'}{(2\pi)^3}e^{i\mathbf k'\cdot(\mathbf r-\mathbf r')}\right]}_{=\;\delta(\mathbf r-\mathbf r')}
\;\int \frac{d^3\mathbf q}{(2\pi)^{3/2}}\,C_0(\mathbf q)\widetilde V(\mathbf q)\,e^{-i\mathbf q\cdot\mathbf r'}\nonumber\\
&=\psi(\mathbf r)\int \frac{d^3\mathbf q}{(2\pi)^{3/2}}\,C_0(\mathbf q)\widetilde V(\mathbf q)\,e^{-i\mathbf q\cdot\mathbf r}.
\end{align}
Renaming $\mathbf q\to-\mathbf q$ gives
\begin{equation}
\big[\mathcal F^{-1}C_0\widetilde V\,\mathcal F\,\psi\big](\mathbf r)
=\psi(\mathbf r)\int \frac{d^3\mathbf q}{(2\pi)^{3/2}}\,C_0(\mathbf q)\,\widetilde V(\mathbf q)\,e^{+i\mathbf q\cdot\mathbf r}
\;\equiv\;V_{\text{eff}}(\mathbf r)\,\psi(\mathbf r),
\end{equation}
with the effective potential
\begin{equation}
V_{\text{eff}}(\mathbf r)=\int \frac{d^3\mathbf q}{(2\pi)^{3/2}}\,C_0(\mathbf q)\,\widetilde V(\mathbf q)\,e^{i\mathbf q\cdot\mathbf r}.
\end{equation}

Therefore, the real-space equation is
\begin{equation}
\big[\mathcal F^{-1}\epsilon\,\mathcal F\,\psi\big](\mathbf r)+V_{\text{eff}}(\mathbf r)\,\psi(\mathbf r)=E\,\psi(\mathbf r),
\end{equation}
and, finally, in compact operator notation (matrix notation, omitting explicit momentum arguments on $\epsilon$),
\begin{equation}
\boxed{\,(\,\mathcal F^{-1}\,\epsilon\,\mathcal F \;+\; V_{\text{eff}}\,)\,\psi \;=\; E\,\psi\, .}
\end{equation}

Here, the kinetic operator remains non-local due to the band dispersion $\epsilon(\mathbf{k})$, while all complexity associated with the Bloch functions has been abstracted into the effective potential kernel \(C_0(\mathbf{q})\). This allows the valley-induced fast oscillations to be captured in the wavefunction $\psi(\mathbf{r})$, without requiring explicit resolution of the atomistic-scale Bloch modes.

This formulation forms the basis of the generalized envelope function approach used throughout this work.

\section{Deriving the Position-Dependent Valley Hamiltonian}
\label{app:ValleyHamiltonian}

In silicon, valley splitting arises due to coupling between electronic states located near the conduction band minima at \( \pm k_0 \hat{z} \). A convenient complex basis for describing this two-valley subspace is given by
\begin{equation}
\tilde{\chi}_\uparrow(z) = F(x,y,z)\,e^{i k_0 z}, \quad \tilde{\chi}_\downarrow(z) = F(x,y,z)\,e^{-i k_0 z},
\end{equation}
where \( F(x,y,z) \) is a smooth envelope function that varies slowly on the scale of the lattice constant. These states represent the vertical component of the Generalized Envelope Function (GEF) formalism.

In the limit of a slowly varying potential, the GEF envelope \( F \) will be identical to the conventional envelope function defined in \eqref{eq:env_function_theory}. For this reason, we denote both with the same symbol \( F \), implicitly assuming the correspondence between these frameworks in the long-wavelength limit.

The effective Hamiltonian in this valley basis is:
\begin{equation}
\hat{H}_z^{(\uparrow,\downarrow)}(x,y) =
\begin{pmatrix}
0 & |\Delta(x,y)| e^{+i\phi(x,y)} \\
|\Delta(x,y)| e^{-i\phi(x,y)} & 0
\end{pmatrix},
\end{equation}
where \( \phi(x,y) \) is the spatially varying valley phase from atomic-scale interface variations and the zeros on the diagonal are guaranteed by the time-reversal invariance of the one-dimensional vertical problem.

The eigenstates of this matrix define the rotated valley basis:
\begin{equation}
\chi_g = \frac{1}{\sqrt{2}} \left( e^{i\phi/2} \tilde{\chi}_\uparrow - e^{-i\phi/2} \tilde{\chi}_\downarrow \right), \quad
\chi_e = \frac{1}{\sqrt{2}} \left( e^{i\phi/2} \tilde{\chi}_\uparrow + e^{-i\phi/2} \tilde{\chi}_\downarrow \right).
\end{equation}

These can also be expressed in real-valued form:
\begin{equation}
\chi_g(x,y,z) = \sin(k_0 z + \phi(x,y)) F(x,y,z), \quad
\chi_e(x,y,z) = \cos(k_0 z + \phi(x,y)) F(x,y,z),
\end{equation}
clearly showing that \( \phi(x,y) \) shifts the interference pattern between the two valleys.

Alternatively, in the real fixed basis:
\begin{equation}
\tilde{\chi}_+(z) = \cos(k_0 z) F(x,y,z), \quad \tilde{\chi}_-(z) = \sin(k_0 z) F(x,y,z),
\end{equation}
the rotated eigenstates become:
\begin{equation}
\tilde{\chi}_+(x,y,z) = \sin(\phi) \chi_g + \cos(\phi) \chi_e, \quad
\tilde{\chi}_-(x,y,z) = \cos(\phi) \chi_g - \sin(\phi) \chi_e.
\end{equation}

Assuming \( \hat{H}_z \chi_g = \epsilon_g \chi_g \) and \( \hat{H}_z \chi_e = \epsilon_e \chi_e \) where \(\epsilon_e-\epsilon_g=2|\Delta(x,y)|\), the effective vertical Hamiltonian in the real basis becomes:
\begin{equation}
\hat{H}_z^{(+,-)}(x,y) =
\begin{pmatrix}
\epsilon_g \sin^2 \phi + \epsilon_e \cos^2 \phi & (\epsilon_g - \epsilon_e) \cos \phi \sin \phi \\
(\epsilon_g - \epsilon_e) \cos \phi \sin \phi & \epsilon_g \cos^2 \phi + \epsilon_e \sin^2 \phi
\end{pmatrix}
= \epsilon_g + 2|\Delta| \begin{pmatrix}
\cos^2\phi & -\cos\phi\sin\phi \\
-\cos\phi\sin\phi & \sin^2\phi
\end{pmatrix}.
\end{equation}

\section{Valley Phase Extraction}
\label{app:PhaseExtr}

\subsubsection{2D Extraction: Real-Space Projection Method}

In the projected 2D theory, the two lowest vertical basis states are chosen as a fixed valley basis:
\[
\chi_+(z) = \cos(k_0 z)\, F(z), \qquad \chi_-(z) = \sin(k_0 z)\, F(z),
\]

Any vertical wavefunction $\Psi(z)$ within the valley subspace can be expressed as:
\[
\Psi(z) = A\, \chi_+(z) + B\, \chi_-(z) = F(z) \left( A \cos(k_0 z) + B \sin(k_0 z) \right).
\]

The coefficients $A$ and $B$ are obtained through orthogonal projection onto the basis states:
\[
A = \int \chi_+^*(z) \Psi(z) \, dz, \qquad B = \int \chi_-^*(z) \Psi(z) \, dz.
\]

Comparing the spinor representation to the standard cosine form $\Psi(z) = F(z) \cos(k_0 z + \phi/2)$, we can establish the relationship:
\[
A \cos(k_0 z) + B \sin(k_0 z) = |C| \cos(k_0 z + \phi/2),
\]
where $|C|^2 = A^2 + B^2$ and the valley phase is given by:
\[
\frac{\phi}{2} = -\tan^{-1} \left( \frac{B}{A} \right).
\]

This projection method provides a direct route for extracting $\phi/2$ from the computed 2D spinor wavefunction. 

\subsubsection{3D Extraction: Fourier-Based Method}

Assuming the vertical (excited) wavefunction is of the form:
\begin{equation}
\Psi(z) = F(x,y,z) \left[ e^{-i k_0 z} e^{i\phi/2} + e^{i k_0 z} e^{-i\phi/2} \right] = 2 F(x,y,z) \cos(k_0 z + \phi/2),
\end{equation}
we identify the phase \( \phi/2 \) as a relative phase between components oscillating at \( \pm k_0 \). Taking the Fourier transform \( \tilde{\Psi}(k_z) \), this structure implies:
\begin{equation}
\tilde{\Psi}(k_z) \approx \tilde{F}(k_z - k_0)\, e^{i\phi/2} + \tilde{F}(k_z + k_0)\, e^{-i\phi/2}.
\end{equation}

Assuming \( \tilde{F}(k_z) \) is real and symmetric, the complex phase difference between the components near \( \pm k_0 \) directly encodes the valley phase. We extract this by evaluating the ratio of the Fourier amplitudes near \( +k_0 \) and \( -k_0 \):
\begin{equation}
e^{i\phi} = \left\langle \frac{\tilde{\Psi}(k_z \approx +k_0)}{\tilde{\Psi}(k_z \approx -k_0)} \right\rangle,
\end{equation}
\begin{equation}
\Rightarrow \quad \frac{\phi}{2} = \frac{1}{2} \arg\left( \left\langle \frac{\tilde{\Psi}(k_z \approx +k_0)}{\tilde{\Psi}(k_z \approx -k_0)} \right\rangle \right).
\end{equation}

Here, the angle brackets \(\langle \cdot \rangle\) denote a small average over neighborhoods in \( k_z \)-space centered at \( \pm k_0 \), which improves numerical robustness against noise or sampling artifacts.

\section{SiGe Valley Scan Methodology}
\label{apx:sige_valley}

This appendix details the computational methodology used to benchmark the 2D projection method (SQDv) against full 3D simulations (SQD) for valley physics in Si/SiGe quantum dots. The model captures realistic alloy disorder in the SiGe barrier, the vertical potential variation, and the resulting valley-dependent phenomena.

To simulate the effect of a SiGe alloy, we first generate a vertical concentration profile using a smooth sigmoid function \( c_{\text{Ge}}(z) = c_{\text{max}} \cdot \sigma((z - z_c)/w) \), where \( c_{\text{max}} = 30\% \) is the peak Ge concentration, and \( z_c \) and \( w \) determine the center and width of the transition region (see Fig.~\ref{fig:valley_sige_results}). In our study we set $z_c = 0$ and used a $w = 1nm$ (see \ref{fig:valley_sige_results}). At each grid point, the number of Ge atoms is sampled from a binomial distribution based on this local concentration and the grid cell volume. The resulting atomic distribution models realistic disorder while preserving the macroscopic profile.

\begin{figure}[htb]
    \centering
    \includegraphics[width=0.7\textwidth]{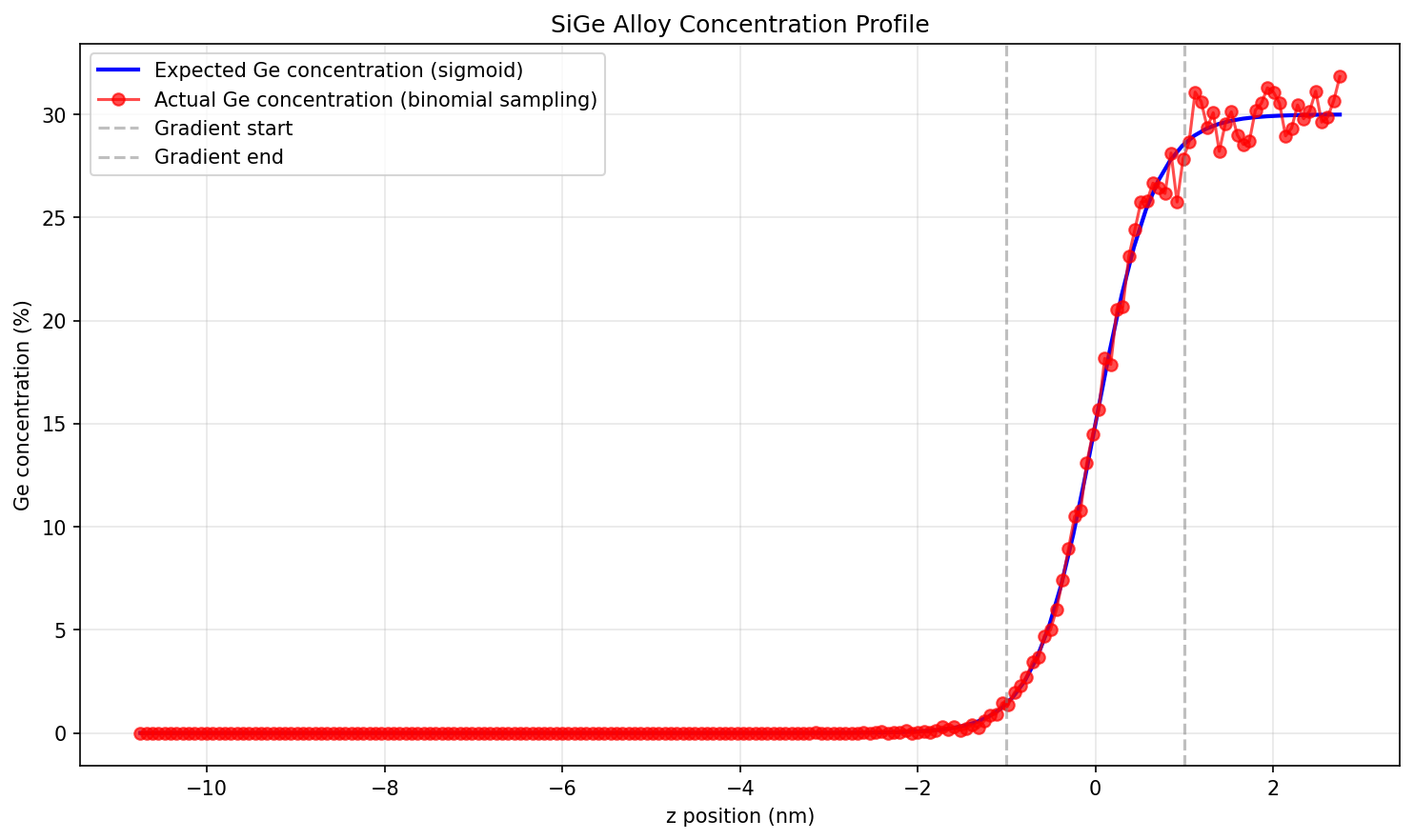}
    \caption{SiGe concentration. The red dots contain the actual germanium concentration obtained via sampling of germanium atoms corresponding to gridpoints along a line in the z-direction. The blue curve illustrates the mean concentration.}
    \label{fig:valley_sige_concentration}
\end{figure}

The local SiGe potential is computed from the sampled Ge content via \( V_{\text{SiGe}}(x,y,z) = V_{\text{max}} \cdot \frac{N_{\text{Ge}}(x,y,z)}{N_{\text{total}} \cdot c_{\text{max}}} \), where \( V_{\text{max}} = 150 \) meV sets the strength of the alloy potential. This is added to a total potential landscape, combining harmonic confinement in the lateral directions and a linear vertical electric field. The overall potential is thus: 
\[
V_{\text{total}}(x,y,z) = \tfrac{1}{2}[\omega_x^2(x - x_0)^2 + \omega_y^2 y^2] - \alpha z + V_{\text{SiGe}}(x,y,z),
\]
with \( \omega_{x,y} = 1.0 \,\text{meV/nm}^2 \), \( \alpha = 5\,\text{meV/nm} \), and \( x_0 \) scanned across the device to emulate lateral dot displacement.


\newpage
\bibliographystyle{unsrt}
\bibliography{references}
\typeout{get arXiv to do 4 passes: Label(s) may have changed. Rerun}

\end{document}